# In Praise of Clausius Entropy: Reassessing the Foundations of Boltzmannian Statistical Mechanics

**Christopher Gregory Weaver***
Assistant Professor of Philosophy, Department of Philosophy[a]
Affiliate Assistant Professor of Physics, Department of Physics[b]
Core Faculty in the Illinois Center for Advanced Studies of the Universe[b]
University of Illinois at Urbana-Champaign
Visiting Fellow, Center for Philosophy of Science[C]
University of Pittsburgh

[*]wgceave9@illinois [dot] edu (or) CGW18@pitt [dot] edu

[a]Department of Philosophy
200 Gregory Hall
810 South Wright ST
MC-468
Urbana, IL 61801

[b]Department of Physics
1110 West Green ST
Urbana, IL 61801

[C]Center for Philosophy of Science
University of Pittsburgh
1117 Cathedral of Learning
4200 Fifth Ave.
Pittsburgh, PA 15260

**Acknowledgments**: I thank Olivier Darrigol and Matthew Stanley for their comments on an earlier draft of this paper. I thank Jochen Bojanowski for a little translation help. I presented a version of the paper at the NY/NJ (Metro Area) Philosophy of Science group meeting at NYU in November of 2019. I'd like to especially thank Barry Loewer, Tim Maudlin, and David Albert for their criticisms at that event. Let me extend special thanks to Tim Maudlin for some helpful correspondence on various issues addressed in this paper. While Professor Maudlin and I still disagree, that correspondence was helpful. I also presented an earlier draft of this work to the Department of Physics at the University of Illinois at Urbana-Champaign in January 2020. I thank many of the physics faculty and graduate students for their questions and objections. A comment from Nigel D. Goldenfeld as well as a challenging question from Michael B. Weissman benefited the final product. Finally, I'd like to express an additional special thanks to Olivier Darrigol with whom I corresponded on various issues in Boltzmann scholarship as well as on numerous points made in this paper. I owe a great debt to him for that enlightening correspondence and for his *Atoms, Mechanics, and Probability* (OUP 2018) from which I learned so much.

**Abstract**: I will argue, *pace* a great many of my contemporaries, that there's something right about Boltzmann's attempt to ground the second law of thermodynamics in a suitably amended deterministic time-reversal invariant classical dynamics, and that in order to appreciate what's right about (what was at least at one time) Boltzmann's explanatory project, one has to fully apprehend the nature of microphysical *causal* structure, time-reversal invariance, and the relationship between Boltzmann entropy and the work of Rudolf Clausius.

"Die Energie der Welt ist constant. Die Entropie der Welt strebt einem Maximum zu."[1]

- Rudolf Clausius (1822-1888)

"For Boltzmann…the probability calculus was primarily a technique for evading paradox; the mechanical approach to gas theory…exemplified by the H-theorem, was always his fundamental tool, the one to which he returned again and again."[2]

- Thomas Kuhn (1922-1996)

"Let us now tum to the second matter in dispute between us. That the majority of students don't understand philosophy doesn't bother me. But can any two people understand philosophical questions? Is there any sense at all in breaking one's head over such questions? Shouldn't the irresistible pressure to philosophize be compared with the nausea caused by migraine headaches? As if something could still struggle its strangled way out, even though nothing is actually there at all?

My opinion about the high, majestic task of philosophy is to make things clear, in order to finally heal mankind from these terrible migraine headaches. Now, I am one who hopes not to make you angry by my forthrightness, but the first duty of philosophy as love of wisdom is complete frankness. Through my study of Schopenhauer, I am learning Greek ways of thinking again, but piecemeal."[3]

- Ludwig Boltzmann's (1844-1906) letter to Franz Brentano (Vienna, January 4th, 1905)

[1] (Clausius 1865, 400).
[2] (Kuhn 1978, 70-71).
[3] (Boltzmann 1995, 125).

## 0 Introduction

With respect to an empirically successful physical theory T, it is believed that one can use T to acquire approximately true descriptions and explanations of phenomena in the world without appreciating T's historical development if one has and uses a universally accepted formulation and partial interpretation of T (*cf.* the remarks in Uffink 2007, 923). In contemporary statistical mechanics (SM) for the classical limit, everyone does what is right in their own eyes (borrowing some wording from Judges 21:25). Unlike contemporary Minkowskian special relativity, or classical Maxwellian electrodynamics, there is no generally agreed upon formulation or approach to non-equilibrium or equilibrium SM. Aside from the (a) Gibbsian approach[4], there are (b) epistemic and information theoretic approaches including some with and some without the Shannon entropy[5], (c) Boltzmannian approaches with and without ergodicity that use the Boltzmann entropy[6], (d) stochastic dynamical approaches that modify the underlying classical microdynamics[7], (e) the Brussels-Austin School[8], and (f) the BBGKY Hierarchy approach, a type of chimera that includes both Gibbsian and Boltzmannian ideas.[9]

Scientific realism is the view that most of the unobservables that are essential to our best physical theories exist and that most property attributions to the self-same unobservables expressed in statements essential to our best physical theories are at least approximately true. Given realism, the multifarious ways of formulating and interpreting SM should not be brushed off as harmless. Each formulation and interpretation of SM recommends a distinctive scientific ontology. For example, some theories provide competing characterizations or interpretations of quantities like entropy. The Boltzmannians claim that thermodynamic entropy is the Boltzmann entropy ($S_B$), an objective property of physical systems whose mathematical representative can change its value over time (*q.v.*, n. 11). For many Gibbsians, $S_G(\rho)$ or the Gibbs entropy is thermodynamic entropy.[10] $S_G(\rho)$ is a constant of motion on the assumed classical Hamiltonian mechanics. It is a time-independent function of a density or probability distribution associated with modal system ensembles. Sometimes interpretations of one and the same formulation of SM can imply different scientific ontologies. For example, different assumptions about the interpretation of probability in one and the same approach recommend non-identical scientific ontologies. One might adopt (a) and yet understand the involved probabilities to be propensities that cause relative frequencies. Or one could remain Gibbsian and yet believe that probabilities in the theory just are frequencies solely (*i.e*., propensities are removed from the interpretation of the formulation of (a)). Who is right? Some will shout the answer: "That theory or approach which enjoys the most empirical success is the theory that is closest to the truth!" (remember that I am assuming realism).

---

[4] For which see the standard presentation in (Landau and Lifshitz 2005); and *cf.* the discussions in (Frigg and Werndl 2019); (Peliti 2011, 55-88, but especially 84-86); (Thorne and Blandford 2017, 155-218, 246-282) and (Uffink 2007, 992-1005).

[5] See (Jaynes 1983, 4-38, 210-336); (Ladyman, Presnell and Short 2008); and while (Tolman 1979, 59-70) does seem to be a proponent of an epistemic approach, entropy does not appear to be epistemically or information theoretically interpreted at ibid., 561.

[6] See (Albert 2000); (Albert 2015); (Callender 2011); (Goldstein and Lebowitz 2004); (Goldstein, Tumulka and Zanghì 2016); (S. Goldstein, Huse, et al. 2019); (S. Goldstein, et al. 2019); (Lebowitz and Maes 2003); (Loewer 2008) and (Penrose 2012, 11-79).

[7] See the results and literature discussed and cited in (Seifert 2012, especially section 2). See also the interesting work being done at (SISSA).

[8] See (Prigogine 2003) and (Prigogine and Stengers 1984). Also see the discussions of this approach in (Batterman 1991); (Bishop 2004); (Bricmont 1996), (Earley 2006), and (Karakostas 1996).

[9] See (Cercignani 1998, 261-263); (Cercignani, Illner, and Pulvirenti 1994); (Cercignani, Gerasimenko and Petrina 1997). Also see the discussions of this approach in (Uffink 2007, 1034-1038) and (Uhlenbeck and Ford 1963, 118-138).

[10] Suppose the real world statistical mechanical system of *n*-material points (*SYS*) of interest is in microstate $x$ at an initial time $t_0$. Call this microstate of *SYS*, $x_0$. In both Gibbsian and Boltzmannian SM, $x_0$ is represented by a point on a $6N$-dimensional phase space $\Gamma$ over which is defined the standard Lebesgue measure $\mu$. $N$ is the number of molecular or particle-constituents. The point itself represents the positions and momenta of the micro-constituents of the system. $\Gamma$ will have coarse grained regions with volumes in the phase space. One could also understand these regions as subsets of $\Gamma$ over which one can define a $\sigma$-algebra (as in the explication in Frigg and Werndl 2019). The evolution of *SYS* from $x_0$ to some other microstate at a later time, is given by an evolution function $\phi_t$, a measure-preserving flow or phase space orbit that is determined by solutions to the equations of motion (Hamilton's equations).

Suppose the exclaimed answer is correct and that (a)-(f) could somehow be empirically distinguished. Let us further suppose that in point of fact, the deliverances of experimentation and scientific observation privilege (c) *the Boltzmannian approach*. Like many modern promulgators of (a)-(b) and (d)-(f), defenders of standard Boltzmannian SM (BSM) claim to be in the possession of a *unique* ideological solidarity with those fathers of modern kinetic theory and statistical mechanics that are James Clerk Maxwell (1831-1879), Ludwig Eduard Boltzmann (1844-1906), and Josiah Willard Gibbs (1839-1903).[11] In light of the dizzying array of approaches, proponents

---

The later microstate of *SYS* as fixed by the evolution function is represented by $\phi_t(x_1)$, and the evolution from $x_0$ to that subsequent microstate of *SYS* is itself represented by a curve on Γ. If we were to imagine the microstate of *SYS* that is *x* traveling on the curve from $x_0$ to $x_1$, the volumes of the coarse-grained regions would remain the same, since the two approaches in view assume Liouville's theorem. For details on Liouville's theorem see (Taylor 2005, 543-546).

Let an ensemble be a hypothetical infinite collection of non-interacting systems with the same structure as *SYS*, although every member of the ensemble represents a different physical state of *SYS*. The actual state of *SYS* is still given by a point *x* on Γ, but the ensemble itself can be represented as a cloud of phase space points. A phase space orbit of the cloud from one coarse-grained region to another represents the evolution of the ensemble over time.

That $x_t$ is located in a particular region R of Γ at a time *t* has a probability $p_t$(R),

(Eq. 1.n10): $p_t(R) = \int_R \rho(x,t)\, dx$

where $\rho$ is the ensemble, although it can be understood as supplying a probability, *viz*., the probability of the microstate of the system residing in a particular region of the phase space (or $p_t(R)$).

We can now use the probability density $\rho(x,t)$ to define the Gibbs entropy $S_G(\rho)$ as follows,

(Eq. 2.n10): $S_G(\rho) = \int \rho \ln \rho\, dx$

Given that *SYS* is in thermal equilibrium and that it is exemplifying a macroscopic physical quantity P, we can connect P with *f*, the latter having a phase average $\langle f \rangle$.

(Eq. 3.n10): $\langle f \rangle = \int_x f(x)\rho(x,t)\, dx$

We can now say that the value of variable *f* connected with quantity P exemplified by *SYS* in thermal equilibrium will be its phase average $\langle f \rangle$. The variable *f* here is then a macroscopic variable. My explication leans on the sources cited in note 4, and I especially lean on (Frigg and Werndl 2019, 426-429); *cf.* (Goldstein and Lebowitz 2004, 64-65); and (Tolman 1979); *cf.* (Taylor 2005)).

[11] *E.g*., see the remarks in (Albert 2000, 76 n. 5 who emphatically references Gibbs), (S. Goldstein 2001, 39-40) as well as those in (Goldstein and Lebowitz 2004, sect. 2) where the authors attempt to connect what is sometimes called Boltzmann's combinatorial entropy formula (*i.e*., $S_B(X) = k \log vol\, \Gamma(X)$ or the Boltzmann entropy of macrostate *X* of a physical system *SYS* equals the Boltzmann constant multiplied by the logarithm of the volume of the phase space region representative of *X*) with what they identify as Clausius entropy. However, their discussion of Clausius entropy leaves much to be desired. They quote Rudolf Clausius's (1822-1888) statement of the second law (*q.v.,* my n. 1 above) and then discuss textbook presentations of thermodynamic entropy. There is an attempt to show

of BSM have tried to distinguish their perspective, not just by pointing to their theory's empirical success, but also by telling a story (the **Standard Story**) about how Boltzmann came to affirm a combinatorial characterization of entropy (*q.v*., n. 11) and a statistical statement of the second law of thermodynamics (*q.v*., **appendix 1**).

According to the **Standard Story**, from 1866[12] to 1877[13], Boltzmann hoped to provide a purely mechanical justification of the second law, eventually (as of 1872[14]) relying upon his famous minimum theorem (later called the H-theorem). However, Boltzmann's efforts were met by the famous reversibility objection articulated by his colleague Johann Josef Loschmidt (1821-1895) in 1876.[15] In 1877[16], Boltzmann repented and turned to his combinatorial arguments wherein was featured combinatorial entropy (*qv*., n. 11) and a statistical understanding of the second law. He subsequently (to quote one renowned historian of physics) turned "his attention to other matters, returning…only occasionally, to add a footnote or two to his earlier expositions, or to comment on some other physicist's discussion…" (M. J. Klein, Ehrenfest 1970, 108).[17]

---

in (ibid.) that the Boltzmannian approach defended there has a distinguished pedigree because its ideas go back to Boltzmann and some of Boltzmann's ideas about entropy align with or at least can capture some facets of Clausius's work on thermodynamic entropy.

The aforementioned equation for the Boltzmann entropy was first proffered by Planck, *not* Boltzmann (Kragh 1999, 61). Naturally enough, it was also Planck who first introduced *k* ("Boltzmann's constant") into physics.

[12] See (Boltzmann, On the Mechanical Significance [Meaning] of the Second Law of Heat Theory 1866).

[13] See (Boltzmann, Comment on Some Problems in Mechanical Heat Theory 1877) wherein Boltzmann responds to Loschmidt.

[14] See (Boltzmann, Further Studies on the Thermal Equilibrium of Gas Molecules 1872).

[15] See (Loschmidt 1876, 139). It was Boltzmann who in (Boltzmann, Comment on Some Problems in Mechanical Heat Theory 1877), adjusted Loschmidt's reasoning, turning it into what is now commonly called Loschmidt's paradox (and here I'm agreeing with (Darrigol 2018, 195)).

[16] (Boltzmann, On the Relation between the Second Law and Probability Calculus 1877).

[17] Here is Klein's more complete depiction (as found elsewhere) of what I am calling the Standard Story:

> "It was Boltzmann who showed how irreversible behavior could be explained and who obtained an expression for the entropy in terms of the molecular distribution function. Under the pressure of Josef Loschmidt's criticism of his H-theorem of 1872, Boltzmann constructed a fully statistical explanation of the second law, in which irreversibilty [*sic*.] was to be understood as the normal evolution of a system into the most probable state, that is, the most probable molecular distribution allowed by its circumstances.
>
> Boltzmann reached this fully statistical interpretation of the second law of thermodynamics in 1877. He evidently believed that the problem was settled, that he had explained the essential features of the second law, and he turned his attention to other matters. His later discussions of this problem, in the 90's, were undertaken only in response to new criticisms, *and always consisted of elaborations and more careful restatements of his statistical point of view*." (M.J. Klein, Mechanical 1973, 63 emphasis mine)

Klein would add that Boltzmann liked Hermann von Helmholtz's (1821-1894) attempt to provide a mechanical analogy for thermodynamics in the 1880s. Boltzmann explored the analogy himself. Klein goes so far as to suggest that Boltzmann accepted Helmholtz's analogy suitably amended (ibid., 70).

The shift to analogical considerations in Boltzmann's thought is not typically part of the **Standard Story** in the work of contemporary Boltzmannians (*e.g*., (Albert 2000); (Albert 2015); *cf*. the not so contemporary (Ehrenfest and Ehrenfest 1990)). See also (Sklar 1993, 32-44), although Sklar seems to maintain that Boltzmann's combinatorial view was an attempted probabilistic interpretation of the *H*-theorem (ibid., 41). That reading is suspect because the functional *H* (or -*H*) in the H-theorem "does not [always] correspond to the Boltzmann entropy" in Boltzmann's combinatorial work (S. Goldstein, et al. 2019, 28). Sklar does express doubts about acquiring a definitive interpretation of the original literature at (Sklar 1993, 37). Other proponents of at least key parts of the **Standard Story** include Tim Maudlin, as his view on this matter was presented at the 2019 *Foundations of Physics Workshop: A Celebration of*

In what follows, I challenge the Standard Story while also providing the beginnings of a Boltzmannian approach that stands in true solidarity with Boltzmann's corpus. Like Hans Christian Ørsted's (1777-1851) reason for seeking a discovery of the interaction between electricity and magnetism (his Naturphilosophie)[18], or one of Maxwell's reasons for preferring a field ontology in electrodynamics (*viz*., that causes must be spatiotemporally local)[19], or one of Albert Einstein's (1879-1955) reasons for preferring the Lorentzian spacetime of general relativity to the Minkowski spacetime of special relativity (*viz*., the action-reaction principle)[20], my Boltzmannian outlook is motivated by a *metaphysical* thesis, a thesis that is friendly to (what I will show in **sect. 4.1** was) Boltzmann's aim to mechanically explain[21] the process of entropic increase:

---

*David Albert's Birthday* at Columbia University under the title "S = k ln (B(W)): Boltzmann Entropy, the Second Law and the Architecture of Hell". Brown et. al. (2009, 185, 187) and Uffink (2007, 967) affirm that part of the Standard Story which emphasizes an abandonment of the H-theorem (understood as an exceptionless and deterministic understanding of thermodynamics) in the face of Loschmidt's reversibility objection. They affirm that Boltzmann replaced the H-theorem and its mechanical approach to justifying the second law with a probabilistic or statistical outlook. This can also be seen in the work of Brown and Myrvold in (2008). They remarked,

> "…in his 1895 reply to Culverwell *et. al*., Boltzmann is reiterating the probabilistic position he adopted in his first 1877 paper in response to Loschmidt's objection to the original form of the H-theorem…from 1877…[a specific] process of equilibration becomes for Boltzmann merely probable…The change in thinking is particularly evident in the treatment of the homogeneity of the gas. For Boltzmann, in 1872, once this condition is achieved it is permanent. But in 1877, he flatly denies such permanence for arbitrary initial states. *The understanding of irreversibility has taken on a new form*, despite some very misleading remarks by Boltzmann to the contrary. The significance of this shift of reasoning…cannot be overstressed…" (ibid., 26-27 emphasis in the original; these authors include a section (8.2) entitled, "Post-*H*-theorem Boltzmann: Probability reigns" (ibid., 29))

Dürr and Teufel stated that Loschmidt's,

> "reversibility objection…led Boltzmann to recognize that his famous *H*-theorem…which in its first publication claimed irreversible behavior for all initial conditions, was only true for typical initial conditions. Because, as Boltzmann immediately responded, [they have in mind (Boltzmann, On the Relation between the Second Law and Probability Calculus 1877)] these bad initial conditions are really very special, more atypical than necessary." (2009, 87)

Ben-Menahem and Hemmo have written,

> "…Boltzmann's H-theorem turned out to be inconsistent with the fundamental time-symmetric principles of mechanics. This was the thrust of the reversibility objection raised by Loschmidt…It is at this juncture that probability came to play an essential role in physics. In the face of the reversibility objections, Boltzmann concluded that his H-theorem must be interpreted probabilistically." (2012, 5-6)

We can add to this list a Nobel Laureate, (Segrè 1984, 244-245). All these thinkers seem to be under the heavy influence of (Ehrenfest and Ehrenfest 1990). Indeed, some of them note the influence (*e.g*., Ben-Menahem and Hemmo state that they "essentially follow the Ehrenfest and Ehrenfest reconstruction of Boltzmann's ideas in a very schematic way" (2012, 5. n. 6)). On this influence of the Ehrenfests, see (Badino 2011, 354). My turn away from the **Standard Story** follows (*with important differences and departures*) (Badino 2011), (Kuhn 1978), and (von Plato 1994).

[18] See (Friedman 2007).

[19] See (Heimann 1970, 173-174).

[20] See (Brown and Lehmkuhl 2017).

[21] There are important contemporary studies of mechanism and mechanistic explanation in the work of (Glennan 2017) and within the contributions to (Glennan and Illari 2018). Some of the views expressed in the

(**Causal Collisions** (**CC**)): Within the collisions that are quantified over by the hypothesis of molecular chaos (**HMC**) (**sect. 7.1**) and that produce entropic increase thereby making true the Boltzmann equation (**sect. 3**) and H-theorem (**sect. 4**) are instances of an obtaining fundamental causal relation that is formally and *temporally* asymmetric. Particular instances of this fundamental relation in evolutions of thermodynamic systems necessitate one-sided chaos and produce the velocity correlations referenced by the (**HMC**).

I will detail precisely how (**CC**) does a surprising amount of explanatory work (it earns its keep) primarily by arguing that it enables one to respond to the reversibility objection *without* having to endorse Boltzmann's combinatorial arguments.

## 1 The Maxwell Distribution

Let's travel back in time to the year 1859 at a meeting of the British Association for the Advancement of Science in Aberdeen Scotland. Motivated by the 1857 Adams Prize that was announced in March of 1855, James Clerk Maxwell has just completed his studies on the rings of Saturn.[22] Those studies involved the utilization of probabilistic reasoning in physics ((Harman 1998, 128); (Segrè 1984, 168)) as well as reflection upon complex systems of colliding bodies (Harman 1990, 25). What is more, that reasoning and reflection primed Maxwell for the development of contributions to, and investigations of the early kinetic theory of gases.[23]

Maxwell's presentation at the aforementioned 1859 meeting is entitled "Illustrations of the Dynamical Theory of Gases" and it will be published in two parts in *The Philosophical Magazine* a year later.[24] Maxwell's exposition of proposition IV (found later in (Maxwell, Part 1 1860, 22-24)) includes a heuristic argument for a particular hypothesis concerning the velocity distribution for gas molecules understood as elastic spheres composing a gas at uniform pressure. A velocity distribution for a gas system is a quantitative description of the molecular velocities enjoyed by the constituents of the gas at particular temperatures. Velocity distributions can give one both the

---

aforementioned sources can be used to help further develop the sense in which my Boltzmannian approach provides a mechanistic explanation of entropic increase.

[22] See (Maxwell 1990 vol. 1, 438-479) and (Brush, Everitt, and Garber 1983). The Adams Prize was named after John Couch Adams (1819-1892) who predicted the existence of Neptune in 1845. Of course, Maxwell won the 1857 Adams Prize, but it appears that his essay was the only one submitted for it. See P.M. Harman's note (2) in (Maxwell 1990 vol. 1, 438-439). See also (Maxwell 1859); (Maxwell 1983 2).

[23] For an introduction to the ten tenets of *modern* kinetic theory and some thoughts about how Maxwell contributed to that modern understanding, see (Holton and Brush 2006, 311-315).

[24] See (Maxwell, Part 1 1860) and (Maxwell, Part 2 1860). These two papers are misread by prominent contemporary philosophers of physics. For example, Frigg and Werndl (2011, 123) state that Maxwell assumes in his 1860 work that the constituents of gas systems do not interact. They state Maxwell shows how in equilibrium, the relevant gas types are described by the Maxwell-Boltzmann distribution (*q.v.*, **sect. 2**). I don't know how Maxwell could have shown this. The Maxwell-Boltzmann distribution isn't introduced until 1868. Boltzmann doesn't propose a distribution law until then. Second, the gas constituents do interact through contacts or impacts in collisions. Even hard spheres can exert impact forces upon one another although they do not attract or exert repulsive forces upon one another. Furthermore, the subsystems would interact even if Maxwell restricted his attention to ideal gases. I make this last point because many seem to believe that ideal gas molecules don't interact at all (ibid., 127; Frigg 2008, 119). This is false. See footnote 191 below.

average and most probable molecular speeds of constituents of a gas system at various temperatures. From knowledge of average molecular speeds multifarious phenomenological properties can be inferred.

If we glide forward in time to 1867, we'll find Maxwell at his family estate (Glenlair House) just before he'd become the first Cavendish Professor of Physics at Cambridge. There Maxwell publishes "On the Dynamical Theory of Gases"[25] in the Royal Society's *Philosophical Transactions* after admitting the existence of problems with his 1860 theory of gas diffusion revealed in criticisms from Clausius in (Clausius, Conduction 1862).[26] That paper sharpens some of his 1860 argumentation in that several of the assumptions of the 1860 work are abandoned in favor of more realistic assumptions, although both the 1860 and 1867 projects maintain the spirit of some earlier correspondence between Maxwell and Sir George Gabriel Stokes (1819-1903).[27]

Contrary to the reigning paradigm of thought at the time (especially in the work of Clausius[28]), Maxwell hypothesized that collisions between gas molecules over time do not produce the same or close to the same velocities for every constituent molecule of a gas system, although the molecular kinetic energies are caused by those collisions to equal or closely approach the same value.[29] Rather, over time, collisions produce a distribution of speeds or velocities. *All* velocities and positions of the molecular constituents consistent with the conservation laws and the system's total energy are assumed to be nomologically possible as the system evolves.

For Maxwell, collisions are causal phenomena, as are the processes of physical systems that lead to them.[30] The reason why Maxwell believes collisions are causal is because within such processes forces act and those forces are causes.[31] Maxwell includes in the titles of the 1860 and 1867 projects the term 'dynamical'. This is purposeful. As in his celebrated paper "A Dynamical Theory of the Electromagnetic Field" published in 1865[32], Maxwell's approach is dynamical because he's trying to account for the motions of bodies by appeal to causal forces, except in the case of gas systems he does not involve the causal influences of fields. In December of 1866, Maxwell wrote to Stokes as follows: "I therefore call the theory a dynamical theory because it considers the motions of bodies as *produced* by certain forces".[33]

---

[25] See (Maxwell 1867).

[26] See also (Smith 1998, 245, 346 n. 22).

[27] See (Maxwell 1907).

[28] See (Brush 1999, 22).

[29] He said, "my particles have not all the same velocity, but the velocities are distributed according to the same formula as the errors are distributed in the theory of least squares." (Maxwell 1907, 10); (Maxwell vol. 1 1990, 610); (Brush, vol. 1 1976, 233).

For the point regarding kinetic energy, see (Darrigol 2018, 301-310).

[30] With respect to collisions, see (Maxwell vol. 1 1990, 380, 405). With respect to velocities, note the draft comments at (Maxwell vol. 1 1990, 135), where he says, "[t]he external cause which sustains the motion of agitation in the case of Saturn's rings is the different velocities…".

[31] "When the objects are mechanical, or are considered in a mechanical point of view, the causes are still more strictly defined, and are called *forces*." (Maxwell vol. 1 1990, 378 emphasis mine) Harman adds in note (6) of ibid., "Compare Whewell's view that the idea of cause construed as force is the 'fundamental idea' of mechanics", subsequently citing (Whewell 1840, 177-254, 437-494) *inter alia*. Whewell influenced Maxwell's work as is evidenced by Maxwell's "Cambridge kinematical research" approach in (Maxwell 1856), quoting (Smith 1998, 305).

In Maxwell's 1873 demonstration of the generalized Maxwell distribution (*i.e.*, the Maxwell-Boltzmann distribution discussed in **sect. 2** below) Maxwell very clearly invokes forces understood as causal mechanisms that influence the motions of gas molecules and act on systems (Maxwell 1873, 537-538).

[32] (Maxwell 1865).

[33] (Maxwell vol. 2: part 2 2009, 291) emphasis mine. After quoting this precise passage, Harman (1998, 127) adds "[t]his defines the dynamical basis of his theory of gases." There is, of course, a sense in which (as Maxwell

Maxwell is propelled into his particular way of studying gas systems by reading Clausius's 1859 memoir "On the Mean Length of the Paths Described by the Separate Molecules of Gaseous Bodies".[34] He learns that the key to discerning the properties of gas systems is to look to *collisions* of the constituents of those molecules, which were (as with Maxwell) causal phenomena in the mind and work of Clausius. Clausius believed that around each gas constituent (or a center of gravity) is a "*sphere of action*"[35] determined by the capacity of such constituents to "drive" one another "asunder" (*i.e.*, to repel one another).[36] When a constituent α approaches another constituent β thereby entering β's domain of repulsive influence or sphere of action, a rebounding effect results from a repulsive force, and both α and β (because of Newton's third law of motion) change their velocities (given appropriate inertial masses). The acting repulsive force is causal in that it produces its "effects…at very small distances".[37] In a manner very much dependent upon Clausius, Maxwell's 1867 work maintained that gas systems attain velocity distributions indicative of thermal equilibrium (*q.v.*, equations (0) and (1) below) because of the collisions of their constituents, where again, the collisions were understood by Maxwell to be the causal mechanisms that produce velocity changes.[38] This is all encoded in the underlying mathematics.[39]

---

says) we abandon something like mechanical or dynamical descriptions of physical evolutions when we revert to statistical methods (Maxwell 1891, 339), but that is only because we invoke statistical methods due to our inability to "follow every motion by the calculus." (ibid.) Following every motion by the calculus is what Maxwell calls "the strict dynamical method" (ibid.).

[34] (Clausius 1859). We can judge that Maxwell learned from Clausius in the way I'm suggesting on the basis of correspondence between Maxwell and Stokes dated May 30, 1859 (Maxwell vol.1 1990, 606-611).

[35] (Clausius 1859, 84).

[36] (Clausius 1859, 82-83).

[37] (Clausius 1859, 84). Besides the 'sphere of action' and 'effects' talk, Clausius also uses terms like 'influence'. The "molecular forces are of influence in sensibly altering the motion of the molecule" (ibid., 82). My points are not evaded by resorting to the original German publication of 1858.

[38] (Porter 1986, 126). Maxwell thought that the Maxwell distribution is stable under collisions given that the number of a particular set of collisions $d\nu$ equals the number of reciprocal collisions $d\nu'$. Collisions have pre and post-collision velocities. If there's a binary collision—Maxwell restricted his reasoning to binary collisions—with pre-collision velocities $\mathbf{v}_1$ and $\mathbf{v}_2$ and post collisions velocities $\mathbf{u}_1$ and $\mathbf{u}_2$, then its reciprocal is the binary collision with pre-collision velocities $\mathbf{u}_1$ and $\mathbf{u}_2$ and post-collision velocities $\mathbf{v}_1$ and $\mathbf{v}_2$. It is not a trivial matter whether there are such reciprocal collisions for any set of existing collisions (*q.v.*, the discussion of Lorentz and Boltzmann in **sect. 4** below). Later on, Maxwell asserted that there are such reciprocal collisions if the colliding objects are perfectly elastic (or perhaps point-like) molecules acting through central forces (Maxwell 1873, 537).

[39] Following Darrigol's (2018, 81-83) reading of Maxwell, restrict the mind's attention to a gas system S whose point-like molecules influence each other through central forces that only engage in binary elastic collisions. Consider that for Maxwell, there are a number of binary collisions $d\nu$ belonging to a particular collision-type $\sigma$. Suppose that the two colliding molecules are $M_1$ and $M_2$ that had pre-collision velocities $\mathbf{v}_1$ and $\mathbf{v}_2$ (respectively) and that took on post-collision velocities $\mathbf{u}_1$ and $\mathbf{u}_2$ (respectively). For Maxwell, whether a collision is of the $\sigma$-type depends upon collision parameters that are the azimuthal angle and the impact parameter (Maxwell 1867, 56-57). The latter consists of the two paths the colliding molecules would have traveled were they to fail to interact with one another (*in the center-of-mass reference frame*). The azimuthal is the angle that fixes the plane upon which sits the post-collision trajectories of both molecules. Let $f(\mathbf{v})\mathrm{d}^3\upsilon$ give the number of molecules per unit volume that enjoy velocities within the $\mathrm{d}^3\upsilon$ range about velocity $\mathbf{v}$. And let $q$ represent a property of any molecule in S, *e.g.*, kinetic energy or inertial mass. Collisions can and do change the total value of $q$ within a specific velocity element $\mathrm{d}^3\upsilon_1$. That change wrought by collisions is encoded by the equation (Darrigol 2018, 82):

(Eq. 1.n38): $\delta[q_1 f(\mathbf{v}_1)\mathrm{d}^3\upsilon_1] = \int_{\mathbf{v}_2\sigma}(q_1' - q_1)\, d\nu$

If we were to suppose that molecule $M_1$ (or any molecule for that matter) enjoys a velocity within $\mathrm{d}^3\upsilon_1$, and that $M_1$ collides with $M_2$ (a distinct molecule that enjoys pre-collision velocity $\mathbf{v}_2$), that collision will produce a variation or Maxwell provided a quantitative statement of his velocity distribution. That is to say, he wrote down an equation for $f(\mathbf{v})$ or the average number of molecular constituents in a gas that enjoy a velocity between two limits ($\mathbf{v}$ and $\mathbf{v} + d^3\mathbf{v}$) subsequent to a great many collisions between similar gas constituents (Maxwell, Part 1 1860, 22). The 1867 statement of $f(\mathbf{v})$ takes the general form $f(\mathbf{v}) = \alpha e^{-\beta u^2}$(where the velocity $\mathbf{v}$ is a three-vector with three Cartesian components $v_x$, $v_y$, and $v_z$, the distribution function $f(\mathbf{v})$ is isotropic[40], α and β are constants, *e* is Euler's number (the base of natural logarithms) such that $e \approx 2.71828$, and *u* is mean velocity). Or more precisely,

(0) **Maxwell's Distribution Law (Vector Notation)**: $f(\mathbf{v}) \propto \mathbf{v}^2 e^{-m\mathbf{v}^2/2kT}$

This function was said to satisfy the relation $f(\mathbf{v}_1)f(\mathbf{v}_2) = f(\mathbf{u}_1)f(\mathbf{u}_2)$ for two distinct gas constituents enjoying respective pre-collision velocities $(\mathbf{v}_1)$ and $(\mathbf{v}_2)$, and post-collision velocities $(\mathbf{u}_1)$ and $(\mathbf{u}_2)$.

Maxwell's actual work (which was without modern vector notation) would affirm,

(1) **Maxwell's Distribution Law**: $f(v) = (\frac{N}{\alpha^3\pi^{\frac{3}{2}}})e^{-(\frac{v^2}{\alpha^2})}$(where *N* is the number of gas molecules, and $\alpha^2$ takes a value that is inversely proportional to the gas's absolute temperature)[41]

Equation (1) (or (0)) implies that the distribution function is asymptotically Gaussian. It was understood by Maxwell to give a velocity distribution for the molecules of a gas in thermal equilibrium. He would try to show that his distribution is stable in the sense that collisions among molecules would not disrupt or otherwise change the distribution's applicability to select gases *in equilibrium*. In other words, Maxwell attempted to quantitatively demonstrate that once gas systems achieve equilibrium status they stay there. His attempt failed.[42] Maxwell's failure notwithstanding, important justifications of Maxwell's distribution exist (Brush, vol. 2 1976, 187-188), as do modern versions of his reasoning with suitable fixes (Darrigol 2018, 81-84). We can now claim that for an appropriate restricted set of classical gas systems, Maxwell's distribution is indeed the correct velocity distribution in that it accurately describes the distribution of velocities for those systems in equilibrium. Important experimental confirmation appears in the work of Nobel Laureate Otto Stern (1888-1969) and the 1927 experimentation of John A. Eldridge (b. 1891).[43]

Maxwell attempted to show that his distribution is the only stable distribution under collisions. To do that he used a collision number over some time period of dynamical evolution of the choice gas system (*qq.v.*, n. 38 and n. 39).[44] How Maxwell acquired his collision number thereby attempting to justify his claim regarding stability appears to be mysterious. Numerous commentators have expressed their inability to get past several obscurities and confusions in transmutation of *q* represented by the difference $(q_1' - q_1)$ *that depends* on the collision-type $\sigma$ (which is fixed by the collision parameters) and $M_2$'s pre-collision velocity $\mathbf{v}_2$ (ibid., 81-83).

[40] That is to say, the function is non-directional.

[41] (Brush, vol. 1 1976, 233); (Maxwell 1867, 64). Hendrik A. Lorentz (1853-1928) derived the Maxwell distribution function for monatomic gases and showed its stability under collisions in (Lorentz 1887).

[42] See the discussions in (Brush 1983, 62).

[43] See (Eldridge 1927) and (Stern, Eine direkte Messung 1920); (Stern, Nachtrag zu meiner Abeit 1920), *cf.* (Stern 1946). See the helpful commentary in (Andrews 1928); (Toennies et. al. 2011).

[44] (Maxwell 1867, 58ff.); (Maxwell 1873).

Maxwell's 1867 argumentation.[45] However, everyone seems to agree that his 1860 and 1867 reasoning made use of other assumptions some of which are probabilistic. I have found at least five. I explicate three of them below leaving the last two assumptions about the nature of collisions for **sect. 7**.

(2) The constituents of gas systems are centers of force and can be regarded as what we now call "Maxwell molecules", *i.e*., point-like molecules (for all intents and purposes point-masses) or collections thereof that "move about as a single body"[46] that interact by means of central repulsive forces inversely proportional to the fifth power of the distance between them.[47]
(3) Every direction of particle rebound subsequent to a binary collision is equally probable.[48]
(4) If (3), then both ((a) all three velocity components of any involved velocity have independent probability distributions (and) (b) every displacement direction is as likely as every other).

Notice that assumptions (3) and (4) are at least in part about probabilities.[49]

Are (2)-(4) good assumptions? Leaving aside Maxwell's claim regarding fifth powers, no atomist would baulk at (2). Assumption (4) is proven in (Maxwell 1867). But what about (3)? I shall not appraise it. Maxwell already did. He called it a precarious assumption which he believed put his approach in danger of being altogether unrelated to actual world collisions and interactions.[50] I wish to add only that Maxwell's justification of (3) rested upon the work of Sir John Herschel (1792-1871), specifically Herschel's unsigned 1850 review of Adolphe Quetelet's (1796-1874) work on probability in the 92nd volume of the *Edinburgh Review* (Herschel 1857).[51] This influence is important because we know that Herschel held an epistemic or Bayesian view of probability, maintaining that probabilities are degrees of belief or credences.[52] Herschel's understanding of probability seemed to have rubbed off on Maxwell for one can clearly see an allegiance to an epistemic interpretation of probability in Maxwell's corpus.[53] This should not surprise us. Frequentism was the interpretation of choice in the 20th century, but Bayesianism

[45] See the comments in (Brush, vol. 2 1976); (Brush, vol. 1 1976); (Darrigol 2018, 88); (Everitt 1975); *cf.* (Uffink 2007, 948-952).

[46] (Maxwell 1867, 54).

[47] This assumption did not seem to be essential. Maxwell at times allows for a myriad of possible theories of the underlying microconstituents ((Harman 1998, 126-127); (Maxwell 1867, 54-55); (Smith 1998, 246)) but the actual reasoning does seem to employ (2).

Maxwell claims to have experimentally justified his characterization of the central repulsive forces involved in this assumption (Maxwell 1867, 51).

[48] See (Garber, Brush, and Everitt 1986, 7). This is an assumption of his 1860 work at least. It is still relevant to an assessment of Maxwell's more mature work in 1867. Why? Because in his 1867 paper, Maxwell proves (4), and (4) references (3).

[49] See on these two assumptions (Brush, vol. 1 1976, 186).

[50] (Maxwell 1867, 62 "this assumption may appear precarious"; Garber, Brush, and Everitt 1986, 8).

[51] The connection between Herschel and Maxwell has been established by (Everitt 1975). See also (Brush, vol. 1 1976, 183-189).

[52] Herschel converted John Stuart Mill (1806-1873) to his epistemic view, causing Mill to forsake his objections to Laplace's Bayesian interpretation of probability (Skyrms 2018); (Zabell 2005, 32. n. 18).

[53] (Appleby 2005, 629-630); and see the quotation of Maxwell in (Jeffreys 1983, 1).

reigned supreme in physics during the 19th century (Appleby 2005, 629). These matters will become important later.

## 2 The Maxwell-Boltzmann Distribution

From 1868 to 1871, Boltzmann generalized (1) (*i.e.*, the Maxwell distribution) for gas molecules in such a way that he obtained a velocity distribution for systems of gas molecules with internal and gravitational degrees of freedom (the generalizations eventually captured systems of polyatomic gas molecules[54]).[55] Boltzmann's main result is called the *Maxwell-Boltzmann distribution*. While Maxwell's 1867 distribution took the form: $f(\mathbf{v}) = \alpha e^{-\beta u^2}$ (as noted above), Boltzmann's 1868 distribution took the form: $f(\mathbf{v}) = a e^{-\beta E}$ where $a$ and $\beta$ stand for constants, and $E$ is energy. It's more full content reads,

(5) **Maxwell-Boltzmann Distribution**:
$f(v) = Ae^{-h(\frac{1}{2}mv^2+V[x])}$ (where $A$ is the number of molecules such that that amount normalizes $f$; $h$ is really just $1/kT$ in modern notation, $k$ is Boltzmann's constant)[56] or we could just write: $f(v) = Ae^{-\frac{E}{kT}}$ (where $E$ is total energy).[57]

---

[54] Polyatomic molecules are molecules with more than two atoms that enjoy internal degrees of freedom. They are sometimes described by internal variables that give one their vibrational, rotational, and electronic states (Kremer 2010, 133). Polyatomic molecules therefore have states that are not exhausted by their translational velocities.

[55] (Boltzmann, Studies on the Equilibrium of Live Force Between Moving Material Points 1868); (Boltzmann, On the Thermal Equilibrium Between Polyatomic Gas Molecules 1871); *cf.* (Boltzmann, Further Studies on Thermal Equilibrium among Gas Molecules 1872). See also the comments in the secondary literature at (Jungnickel and McCormmach 1986, 61). According to Darrigol, Boltzmann also realized that one of his generalizations of Maxwell's distribution yields a scientific "approach" that "can be applied to any system of point-atoms whereas Maxwell's original reasoning applies to gases only." (Darrigol 2018, 8).

[56] (Brush, vol. 1 1976, 234); (Segrè 1984, 279). The factor $e^{-h(\frac{1}{2}mv^2+V[x])}$ is called the Boltzmann factor. The Maxwell-Boltzmann distribution in more modern discussions is explicitly dubbed a probability density function (PDF) and more commonly expressed as follows (for ideal gases),

(Eq. 1. n. 56):

$$f(v) = \left(\frac{m}{2\pi KT}\right)^{3/2} 4\pi v^2 e^{[-\frac{m(V_x^2+V_y^2+V_z^2)}{2KT}]}$$

See (Laurendeau 2005, 291)

There are attempts to derive or justify (5) not only in the work of Boltzmann and Maxwell, but also in the work of George Bryan (1864-1928) (who tried to do without certain of Maxwell's assumptions about collision numbers), Kirchhoff (whose argument is similar to Bryan's), Lorentz (whose result is limited), and Max Planck (1858-1947) (whose argument rested on considerations having to do with time-reversal invariance). See (Bryan 1894); (Kirchhoff 1894, 142-148) (see also Kirchhoff 1898); (Lorentz 1887) and (Planck 1895); *cf.* the discussion in (Darrigol 2018, 23-24; 323-327; 358-365) who summarizes Boltzmann's responses to this literature.

In 1894, Boltzmann provided a new derivation of the Maxwell-Boltzmann distribution that did not rely upon any special reasoning or assumptions about collision numbers and that could be extended to polyatomic gas systems (Boltzmann, Application 1894). See the discussion at (Darrigol 2018, 354-355).

[57] Early on (in 1867), Maxwell would say about other types of matter such as polyatomic molecules, that

> "A law of the same general character is probably to be found connecting the temperature of liquid and solid bodies with the energy possessed by their molecules, although our ignorance of the nature of the connexions between the molecules renders it difficult to enunciate the precise form of the law'." (Maxwell 1867, 54)

Due to objections from Francis Guthrie (1831-1898), Maxwell would himself try his hand at deriving this generalized distribution in (Maxwell 1873) and then again in (Maxwell 1879).[58]

Boltzmann tried to prove that any distribution (for the types of gases with which he was concerned) would tend towards (5) (uniqueness), given a change in time, but would later (1898) state in volume two of his *Lectures on Gas Theory* that he could not actually prove its uniqueness (Boltzmann 1964, 313-340).

Although Maxwell did not seem to favor calling $f(v)$ a probability, both he (subsequent to 1867) and Boltzmann interpreted (5) in such a way that it said that the most highly probable energy of a gas molecule takes a value equal to $kT$, where it is understood that molecules could take on energies with a great many other values (consistent with the total energy values and conservation laws) because the likelihood of such energy assignments is never zero. That the distribution given in (5) represents an appropriate gas in equilibrium and that it gives the unique distribution for such a gas system is generally agreed upon by even modern practitioners of what we now call classical statistical mechanics. It is therefore a *bona fide* law of classical theory. Some scholars also maintain that Boltzmann's derivation of (5) is "impeccable" (at least for the non-polyatomic cases), given that the distribution faithfully represents the speeds of molecules in systems at equilibrium (C. Cercignani 1998, 88, and see also 283-286).

## 3 die Fundamentalgleichung

After generalizing the Maxwell distribution so as to obtain the Maxwell-Boltzmann distribution, Boltzmann remarked that "[i]t has thus not yet been demonstrated that whatever the state of the gas may have been at the start, it must always approach the limit discovered by Maxwell."[59] Boltzmann is here concerned with the missing proof of the uniqueness of the distribution function. In (Boltzmann, Further Studies on the Thermal Equilibrium of Gas Molecules 1872), Boltzmann turned to the task of finding an equation (what we would later call the Boltzmann equation) that tracks the evolution of the velocity distribution over time in irreversible processes so as to help reach the missing proof.[60] For cases involving systems with but one species of particle, the Boltzmann equation reads,

(6) **The Boltzmann Equation or Boltzmann's Transport Equation**:
$\frac{\partial f}{\partial t} + \mathbf{a}\left(\frac{\partial f}{\partial \mathbf{v}}\right) + \mathbf{v}\left(\frac{\partial f}{\partial \mathbf{r}}\right) = \frac{\partial f}{\partial t}_{coll.}$, here $f$ is dependent upon time $t$, position $\mathbf{r}$, and velocity $\mathbf{v}$, and it represents the distribution function describing the gas system's state and also how that state evolves, $\mathbf{a}$ represents the accelerations of the particles/molecules between their collisions, and the right-hand side of the equation

[58] See (Guthrie 1873) and the comments in (Brush 1999, 23-24).

[59] "Es ist somit noch nicht bewiesen, daß, wie immer der Zustand des Gases zu Anfang gewesen sein mag, er sich immer dieser von Maxwell gefundenen Grenze nähern muß." *BWA*1, 319-320. (Boltzmann, Further Studies on the Thermal Equilibrium of Gas Molecules 1872); *cf.* (Boltzmann 2003, 266).

Unless I've used the translations of others, all translations from the German into English in this work were assisted by the following resources: (Durrell et. al. 2002); (Strutz 1998); and (Terrell et. al. 2004), plus some software or program assistance by Google Translate and Microsoft Word German language and spell checker software programs (*q.v.*, the acknowledgments).

[60] On the Boltzmann equation, see (C. Cercignani 1988), (Kremer 2010), and (Villani 2002).

or $\frac{\partial f}{\partial t}_{coll.}$ represents *the collision produced* rate of change of the distribution function *f*.[61]

Here is an expression closer to the original work[62],

(7) **Early Boltzmann Equation**:

$$\frac{\partial f}{\partial t} = \int \mathrm{d}\mathbf{v}_2 \int \{f(\mathbf{u}_1)f(\mathbf{u}_2) - f(\mathbf{v}_1)f(\mathbf{v}_2)\}|\mathbf{v}_1 - \mathbf{v}_2| \; d\Omega\sigma(\Omega)$$

where $d\Omega\sigma(\Omega)$ is the differential collision cross section "for a collision in which the relative velocity" after the collision is "in the solid angle $d\Omega$ at $\Omega$ compared to the relative velocity before."[63] The involved integrals are over every possible scattering angle and every possible velocity $\mathbf{v}_2$ of the collision partner. Function *f* is the distribution function, velocities $\mathbf{u}_1$ and $\mathbf{u}_2$ are final (post-collision) velocities, and $\mathbf{v}_1$ and $\mathbf{v}_2$ are initial (pre-collision) velocities.

The Boltzmann equation "completely determines the evolution of the distribution *f* from its initial value".[64] It says how "the distribution" changes "in time under the action of the collisions."[65] And if you can correctly solve for *f*, then with (7) or some form of (6) you'll obtain all that's needed to compute thermodynamic phenomenological properties of the appropriate relevant system. Boltzmann would add that the right side of the equation (7) vanishes such that $\frac{\partial f}{\partial t} = 0$ when the distribution function is Maxwell's, and all other functions tend toward Maxwell's (uniqueness). Boltzmann also affirmed that the velocity distribution will cease to change once it becomes the Maxwell distribution (stability or stationarity).[66]

The literature on the Boltzmann equation is immense. It has grown large for several reasons. First, it has numerous beneficial applications and uses in modern physics.[67] You can use it to figure out transport coefficients (hence "transport equation") for heat conduction, gas interdiffusion, and gas viscosity.[68] And it is utilized in contemporary physics for the study of neutron transport as well as plasma systems. Second, the equation is *not* time-reversal invariant,[69] and the reason why is usually connected to an assumption of the justification of the equation, *viz*., the **HMC** or hypothesis of molecular chaos defined and discussed in **sect. 7** below (Uffink and

[61] (Rennie 2015, 52). In (Boltzmann, Further Studies on the Thermal Equilibrium of Gas Molecules 1872), Boltzmann expressed the equation in terms of integrals that give one how the distribution function (understood as an energy and time-dependent function) changes with time. There are many other versions of this equation in Boltzmann's work. Other forms of expression involve appropriate modifications for various cases in which an external force acts (such as Newtonian gravity) on the evolving system. See (Boltzmann, On the Thermal Equilibrium of Gases on Which External Forces Act 1875).

[62] See (Klein, Ehrenfest 1970, 101).

[63] Ibid.

[64] (Darrigol and Renn 2013, 773).

[65] (Segrè 1984, 243).

[66] Both Maxwell and Boltzmann had argued in favor of this point prior to 1872 (Brush, vol. 1 1976, 237).

[67] (C. Cercignani 1975); (C. Cercignani 1988); (Cercignani and Kremer 2002). See also (Illner and Pulvirenti 1989); (Illner and Shinbrot 1984); (Morgenstern 1954); (Nishida and Imai 1977) for important results on the Boltzmann equation.

[68] With very few mistakes (corrected later by Boltzmann), Maxwell (Maxwell 1867) had already figured out how to try to compute these coefficients without the Boltzmann equation. His efforts used conservation laws which Darrigol says are "implicitly equivalent to the Boltzmann equation" (Darrigol 2018, 12).

[69] See the proof in (Uffink and Valente 2010, 141, 167-168).

Valente 2010). Third, given several assumptions, including the supposition that the gas system under evaluation is dilute and that its constituents are approximated as hard shells, the Boltzmann equation was derived from the time-reversal invariant equations of motion in classical mechanics by Oscar Lanford III (1940-2013).[70] There's some question as to how the irreversibility or asymmetry of the distribution evolution emerges in a way (and this has been demonstrated in (Spohn 1980) and (Spohn 1991)) that avoids the reversibility objection of Loschmidt discussed in **sect. 7** below.[71] I will have more to say about time-reversal invariance and the emergence of irreversibility shortly. For now, let us turn our attention to Boltzmann's minimum theorem (*i.e.*, the H-theorem).

## 4 The H-Theorem[72]

With the Boltzmann equation in hand, Boltzmann thought himself properly equipped for proving the uniqueness of the Maxwell distribution. One can already see how $\frac{\partial f}{\partial t}$ vanishes given that the distribution function is Maxwell's because that function satisfies the relation: $f(\mathbf{v}_1)f(\mathbf{v}_2) = f(\mathbf{u}_1)f(\mathbf{u}_2)$, as I have already noted. Justifying that conditional is not enough to secure uniqueness. One must also show that if $\frac{\partial f}{\partial t}$ vanishes, then the distribution function must be Maxwell's. To acquire the needed demonstration, Boltzmann introduced the time-dependent functional H (not to be confused with the Hamiltonian).[73] He defined that functional in terms of the distribution function *f*.

(8):

$$\mathrm{H} \equiv \int f \log f d\mathbf{v}$$

[70] See (Lanford 1975); (Lanford 1976); (Lanford 1981); and *q.v.*, **Appendix 2**. See (Spohn 1991, 64 and theorem 4.5) for a rigorous statement of the theorem. In some of the relevant literature on Lanford's project, what's shown is that in the Boltzmann-Grad limit and for rarefied gas systems whose molecules are approximated by hard spheres, given smallness of time, that a particular chaos property is exemplified by the choice systems at $t_0$ (and as a consequence temporally propagates for future times), and some other assumptions, one can move from the BBGKY formulation or hierarchy (of equations) to the Boltzmann equation, itself formulated in terms of a hierarchy (the Boltzmann hierarchy). There are proofs which forsake the smallness of time assumption and replace it with a smallness of norm (or smallness of initial data) assumption. See (Cercignani, Illner, and Pulvirenti 1994, 63-84); (Spohn 1991, 48-76 especially p. 76) and the literature cited therein.

Lanford (1981, 75) distinguishes his result from Boltzmann's H-theorem. I'm interested in defending the latter which uses a different chaos property than that which is assumed in work on Lanford's theorem. Boltzmann's chaos property has a **No Mathematics Problem** (defined and solved in **sect. 7** and **Appendix 2** below).

[71] See the discussions in (Cercignani, Illner and Pulvirenti 1994); (Uffink 2007, 1028-1033); and (Uffink and Valente 2010). Given the terminology introduced and defined in **sect. 7**, I maintain that Lanford's project resolves the reversibility objection but does not resolve the **Chaos Asymmetry Problem** or the **No Mathematics Problem**. Uffink and Valente (2010, 160-166) argue for something like the former idea, while Villani (2002, 95-100) agrees with the latter thesis.

[72] My discussion in this section shall pertain to monatomic gases.

[73] Samuel Burbury (1831-1911) introduced H so as to supplant Boltzmann's use of E (S. H. Burbury 1890). Boltzmann would subsequently use H in 1895. Some folks have said that Burbury intended to use η or eta so as to follow Josiah Gibbs's (1839-1903) representation of entropy. That is not true (Brush 2003, 182. first note); (Darrigol 2018, 142. n. 8).

On the assumption that the Boltzmann equation is omnitemporally true, and the assumption that the time and velocity dependent function *f* satisfies the Boltzmann equation, it can be rigorously proven that for any time *t*, the distribution function *f* is Maxwellian, just in case, $\frac{d\mathrm{H}}{dt} = 0$. On the same assumptions (*i.e.*, *f* satisfies the Boltzmann equation and that that equation is omnitemporally true) it can also be proven that,

(9):

$$\frac{d\mathrm{H}}{dt} \leq 0, for\ any\ time\ t$$

But it will turn out that the relevant proofs make use of the hypothesis of molecular chaos (**HMC**) discussed and defined in **sect. 7** below. This was not realized by Boltzmann until sometime after his 1872 and 1875 work.

The conjunction of the above results is called Boltzmann's minimum theorem or H-theorem (Uffink 2007, 965); *cf.* (C. Cercignani 1988, 137-140). The quantity H is a monotonically decreasing function in time unless the velocity distribution is the Maxwell distribution. And so, the theorem helps secure the uniqueness of the Maxwell distribution.[74] As Boltzmann's 1896 summary of the H-theorem in his *Lectures on Gas Theory* stated, "[w]e have shown that the quantity we have called H can only decrease, so that the velocity distribution must necessarily approach Maxwell's more and more closely."[75]

The preceding discussion pertains to Boltzmann's H-theorem for monatomic gases. In his 1872 work, he also tried to prove an H-theorem for polyatomic gases (see also (Boltzmann, On the Thermal Equilibrium of Gases on Which External Forces Act 1875)). That proof did not fare well, as Lorentz found a problem with Boltzmann's derivation.[76] Lorentz notes that part of Boltzmann's derivation of the Boltzmann equation and the H-theorem is a commitment to the existence of reciprocal collisions. Boltzmann appears to assume that if there exists a collision $[A, B] \rightarrow [A', B']$, then there exists an inverse collision (following Lorentz's way of characterizing sets of velocities of colliding molecules) that is $[A', B'] \rightarrow [A, B]$. Lorentz proves that this assumption is false for polyatomic molecules that are non-spherical.[77] He then provides a simplified version of Boltzmann's 1872 proof of the H-theorem for monatomic gases. This streamlined proof is later used by Boltzmann in both his *Lectures on Gas Theory* (Boltzmann 1964), and his 1887 response to Peter Guthrie Tait (1831-1901) entitled *Über einige Fragen der Kinetischen Gastheorie* (On Some Questions about Kinetic Gas Theory). Moreover, modern textbooks often choose to follow Lorentz's proof for the monatomic case when presenting the derivation of the H-theorem for pedagogical purposes (Darrigol 2018, 328-329); (Kox 1990, 599. nn. 38-39).

---

[74] See *BWA*1, 335; (Boltzmann, Further Studies on the Thermal Equilibrium of Gas Molecules 1872). See also (Darrigol and Renn 2013, 773).

[75] (Boltzmann 1964, 55).

[76] See (Brush 1974, 47); (C. Cercignani 1998, 154-155); (Darrigol 2018, 319-327); (Kox 1982); (Kox 1990); and (Lorentz 1887).

[77] (Cercignani and Lampis 1981) argue convincingly that the existence or non-existence of reciprocal collisions depends not so much on the shape of the molecules, but upon the nature of the interactions those molecules are involved in.

Boltzmann would graciously accept Lorentz's criticism and provide a follow-up proof that made use of cycles of collisions as that which drives H-decrease in the polyatomic cases (*q.v.*, n. 76). His maneuver is both unrealistic and embraced by no one, save Lorentz.[78]

So, Boltzmann's attempts at proving an H-theorem for polyatomic gas types had problems. Not even his attempted demonstrations of the H-theorem for the monatomic cases are wholly without problems. A decisive and rigorous proof for the monatomic gas type would have to wait until the 1933 and 1957 work of Torsten Carleman (1892-1949).[79] In addition, Carlo Cercignani (1939-2010) taught us that that demonstration has a cousin yielding an H-theorem for polyatomic gases (C. Cercignani 1998, 96). Both Cercignani and Darrigol have proven an H-theorem for polyatomic gas types (ibid., 287-290); (Darrigol 2018, 493-496).[80] Thus, for both monatomic and polyatomic gas types, we have an H-theorem. How should we interpret it?

### 4.1 Interpreting the H-Theorem: Collisions and Causation

Boltzmann's proposed mechanical explanations of the second law of thermodynamics characterize systems of colliding gas molecules as systems whose constituents *causally* interact. There are four reasons why one should accept this interpretation. First, mechanical explanations of natural phenomena for physicists such as Clausius, Maxwell, and Boltzmann are part of (to quote Christiaan Huygens's (1629-1695) characterization of the mechanical approach, a characterization alive and well during the $19^{th}$ century) "the true Philosophy, in which one conceives the causes of all natural effects in terms of mechanical motions."[81] In other words, a mechanical explanation just is one involving a report on causes that are mechanical motions *inter alia* (*qq.v.*, n. 30, n. 31, and n. 33). In **sect. 1**, I detailed how this approach to mechanical explanation shows up in the work of Clausius and Maxwell. As I shall demonstrate in **sect. 5**, Boltzmann's H-theorem is part of his attempt to mechanically explain the second law. It is therefore highly likely that by Boltzmann's lights, the type of explanation of entropic increase the H-theorem offers is a causal explanation.

---

[78] Interestingly, Boltzmann's Lorentz-inspired argumentation does not make use of the Boltzmann equation. Rather, it "rests on a direct evaluation of the effect of collisions on the value of the H-function." (Darrigol 2018, 327).

[79] (Carleman 1933); (Carleman 1957). The latter was published posthumously. I have not read these papers but was made aware of their contents by the discussion in (C. Cercignani 1998, 96; 273-276). Also see (Villani 2008, 4-8) for a proof sketch.

[80] Darrigol's proof (and compare the proof in (Cercignani and Lampis 1981)) avoids cycles of collisions and discretization techniques. It assumes that the collisions are corresponding collisions. Unlike Boltzmann's Lorentz-inspired proof for polyatomic gas types, it does make use of the Boltzmann equation.

[81] (Huygens 1952, 3). The points I make in this section stand in contrast to the viewpoint adopted in (Badino 2011, 361). There, Badino argues that "Boltzmann…did not draw a clear-cut line between a mechanistic and a probability-based account of a system's approach to equilibrium." (ibid.) The evidence I articulate in the main text that follows shows that Boltzmann thought of mechanistic explanations as special kinds of causal explanations. There is no evidence that he believed causal explanations were provided by his combinatorial approach. As I reveal in **sect. 7.2.1**, the combinatorial approach ignores causal interactions while those ignored instances of causation are central to the H-theorem or mechanistic approach. The latter is more fundamental than the former in Boltzmann's eyes precisely because it says something more directly about the engine of entropic increase, *viz*., causal collisions. That Boltzmann's major influences cut a divide between causal mechanistic explanations and statistical ones is revealed in the remarks of Maxwell's *Theory of Heat*. There, Maxwell said that we abandon mechanical descriptions or explanations of physical evolutions when we appropriate statistical methods (Maxwell 1891, 339).

In several of Boltzmann's lectures, he uses the locution 'mechanical cause'. He does this once in an interesting discussion of medical science. There, Boltzmann speaks as if mechanical explanations are causal explanations (Boltzmann 1974, 133).

Second, in part one of his *Lectures on Gas Theory*, Boltzmann presents a Lorentz-inspired derivation of the H-theorem. In his discussion of value changes of the H-functional, Boltzmann reports that changes in H over a small period of time are "due to…causes" later noting that the changes result from collisions (Boltzmann 1964, 50, and see also 51-52). This suggests that for Boltzmann, the process of entropic increase is a causal process and that collisions for Boltzmann are causal phenomena.

Third, although Boltzmann seems dismissive of metaphysics (he calls metaphysics a "spiritual migraine"[82]) and whilst he views physical hypotheses as pictures or images of the world that are not directly corresponding truths about it, Boltzmann does consistently interpret all forces (and so those forces at work in collisions) causally (Boltzmann 1995, 54). That is to say, he believes that the image of the world supplied by physics depicts the world as a place endowed with causal forces. Commenting on Heinrich Hertz's (1857-1894) 1887 discovery of a form of electromagnetic radiation (*i.e.*, radio waves), Boltzmann causally interpreted the action of the electromagnetic field. He remarked, "…electric and magnetic forces do not act directly at a distance but are caused by changes of state that are propagated from one volume element to the next at the speed of light" (Boltzmann 1995, 84). At the May 29th, 1886 meeting of the Imperial Academy of Science, and so well before gravitation would be reduced to spacetime curvature by Einstein in 1915, Boltzmann causally interpreted the gravitational force (Boltzmann 1974, 17). At the same event, Boltzmann provided a causal characterization of pressure. He said that the molecules involved in thermodynamic systems impinge (or strike) "now more now less strongly, now head on now at an angle" maintaining that when the pressure produced by these impinging molecules is at a point "bigger…we shall at once look for an external cause that moves the molecules to flow preferentially to that point" (ibid., 20).

Some of the strongest evidence for my interpretation of Boltzmann comes from his 1899 Clark University lectures. In them Boltzmann describes the evolution of a gravitating system and says in that context that in general "the cause of motion…we call force", concluding "that at least in this special case acceleration is the decisive feature of force…namely gravity." (Boltzmann 1974, 127-128). Boltzmann added that:

> Kirchhoff rejected the notion that it was the task of science to unravel the true nature of phenomena and to state their first and fundamental metaphysical causes. On the contrary he confined the task of natural science to describing phenomena, a stipulation that he still called a restriction.[83]

Boltzmann would connect Kirchhoff's view to Hertz's in his 1899 Munich lecture: "nobody has yet pointed out that a certain idea [apparently the same idea articulated in the lines just quoted] in Kirchhoff's mechanics if followed to its logical conclusion leads directly to Hertz's ideas" (Boltzmann 1974, 89). Boltzmann would both reject Kirchhoff's view (Boltzmann 1995, 78) and distinguish his understanding of mechanics from that of Heinrich Hertz (1857-1894) by noting that his approach holds on to causal forces, while Hertz's 1899 *Principles of Mechanics in a New Form* abandons them completely.[84] He said that "difficulties arise" for Hertz's approach "as soon as one wants to represent the most ordinary processes of daily experience involving the action of force".[85] Again, forces are at work in the collisions referenced by the Boltzmann equation. Therefore, according to Boltzmann, so too is causation.

[82] (Boltzmann 1995, 144).

[83] (Boltzmann 1995, 78).

[84] You will recall that according to Hertz, hidden masses explain motions. Forces do not. (Hertz 1899, 6, 25-26, 41, 177). See also Boltzmann's summary at (Boltzmann 1974, 90).

Fourth, in some of Boltzmann's notes on natural philosophy put together for a lecture to be given on November 23rd, 1904, Boltzmann said, "[i]t is just its own bad luck that changes in velocity must have a cause", subsequently committing to a view about the *relata* of causation, *i.e.*, that "[a] thing cannot be the cause of a thing, but merely of change."[86] Colliding things produce velocity changes, according to the Boltzmann equation and H-theorem, and so these remarks support my reading. Boltzmann believes that the mechanism of velocity change in the process of entropic increase is a causal mechanism.

We can safely conclude that there's good evidence from Boltzmann's *Lectures on Gas Theory*, Boltzmann's *Lectures on Mechanics*, his personal lecture notes, and his public lecture content that all supports the thesis that Boltzmann endorsed a causal approach to mechanistically explaining the second law.

### 4.2 Interpreting the H-Theorem: Applications and Exceptions

As early as his work in 1872 and 1875 Boltzmann recognized that there could be gas systems that have unique initial conditions such that they do not evolve to the Maxwell distribution. This is because such special systems start out precluding certain velocity and/or position values otherwise consistent with the conservation laws and energy totals. He conjectured that perhaps some constraints on very special systems keep their constituents from realizing all possible values consistent with those laws and totals. As Boltzmann himself put matters when discussing a gas confined to a container, "it is possible that only certain, and not all possible positions and velocities can occur in the course of time (*e.g*., if they were all initially in a line perpendicular to the vessel walls)."[87] It is an assumption of the velocity/energy distribution approach of Maxwell and Boltzmann that every nomologically possible velocity be realizable by gas constituents. I believe Boltzmann was keen enough to realize the connection between special velocity precluding initial conditions and H-theorem inapplicability. My opinion is that as Boltzmann developed a mechanistic explanation of the second law, he knew of possible systems to which the H-theorem could not be applied. My reading is most, and not too, charitable. It entails that it did *not* take the articulation of Loschmidt's reversibility paradox for Boltzmann to come to

---

[85] (Boltzmann 1995, 79). Boltzmann's own treatment of mechanics in (Boltzmann, Vorlesungen über die Principe der Mechanik 1897) gives us more insight into Boltzmann's attitude about Hertzian mechanics, for there Boltzmann would quite clearly disapprove of Hertz's picture (ibid., 1-6; 37-42).

[86] (Boltzmann 1995, 140). Boltzmann's lecture notes on natural philosophy from 1903 to 1906 were brought together by Ilse M. Fasol-Boltzmann. There's some evidence that Boltzmann may not have read these notes verbatim when delivering his lectures. See the comments of John Blackmore at (Boltzmann 1995, 133).

[87] The German reads,

> "Es ist nun möglich, daß nur gewisse, nicht alle möglichen Positionen und Geschwindigkeiten derselben im Verlaufe der Zeit eintreten können (z. B. wenn sie sich zu Anfang alle in einer auf den Gefäßwänden beiderseits senkrechten Geraden befanden)." *BWA*2, 14. (Boltzmann, On the Thermal Equilibrium of Gases on Which External Forces Act 1875) the realization that some monatomic gas systems escaped H-theorem application.[88] The principle of charity is not all that can be said for the proposed interpretation. It explains why (to quote Cercignani) Boltzmann, "when answering" Loschmidt's paradox (discussed in **sect. 7** below):

> did not indicate that he had changed his viewpoint, or that he had deepened his understanding of the subject, as a consequence of the reflections caused by the [reversibility] objection that had been raised against him, but acted as if he were simply re-elaborating his old ideas.[89]

The fact that such a report is correct has perplexed Boltzmann scholars.[90] There exists a challenge to render that report consistent with Boltzmann's judgment that deriving the H-theorem amounts to *rigorously proving* "that whatever the distribution of live force [kinetic energy] may have been at the beginning [initial time], subsequent to a very long time period *it must always* approach that [one] found by Maxwell."[91] The best way to introduce coherence and consistency here is to insist that even in 1872 and 1875 Boltzmann was aware of systems that did not approach Maxwell's distribution on account of the unique initial conditions they enjoyed (agreeing *in part* with (Badino 2011; von Plato 1994) though I am *not* claiming that Boltzmann's H-theorem project was always statistical in the sense that at least Badino seems to have in mind). The necessity of approaching the Maxwell distribution rests upon the assumptions and antecedent of the H-theorem. As I've said several times now, one of the relevant assumptions is that the Boltzmann equation concerns all nomologically possible velocity and position values, *i.e.*, all those that satisfy the laws of conservation (Darrigol 2018, 173 although I disagree with Darrigol's presumption at n. 37). The cases that admit exceptions to the general claim that H always decreases, or that minus-H (where minus-H is proportional to entropy) always increases are cases that *prohibit* some velocity and position values.

#### 4.2.1 Maxwell's (but really Thomson's) Demon and Loschmidt's Exorcism

My reading is controversial. Let me add further lines of support. Boltzmann's proof of the H-theorem appears *after* the articulation of the Maxwell "demon"[92] case, a case well-known for the trouble it produces for any non-statistical and exceptionless statement of the second law. Maxwell discussed it for the first time in a letter to Tait, dated December 11$^{th}$, 1867[93], restating it

---

[88] For a related point see (Darrigol 2018, 171). However, Darrigol adds, "[s]till, there is no reason to think that Boltzmann believes that the H function could fail to decrease in such cases." (ibid.) The excerpt quoted in n. 87 provides the very reason Darrigol believes is missing.

[89] (C. Cercignani 1998 120).

[90] *E.g.*, it seems to have been entertained before in (M.J. Klein, Development 1973).

[91] Emphasis mine. The original German reads as follows:

> "Es ist somit strenge bewiesen, daß, wie immer die Verteilung der lebendigen Kraft zu Anfang der Zeit gewesen sein mag, sie sich nach Verlauf einer sehr langen Zeit immer notwendig der von Maxwell gefundenen nähern muß." *BWA*1, 345. (Boltzmann, Further Studies on the Thermal Equilibrium of Gas Molecules 1872). *Q.v.*, the translation provided by the source at note 105.

[92] Maxwell's discussion referenced a "finite being" (*q.v.*, n. 93) and did not use the term 'demon'. It was William Thomson (Lord Kelvin; 1824-1907) who introduced that notion in his (Thomson 1874, 442, also attributing a definition of the term to Maxwell at the footnote on the same page). Maxwell did not approve of the use of this term (Knott 1911, 215).

[93] (Maxwell, vol. 2: part 1 2009, 328-334, but see specifically 331-332).

in several places including the appendix to his 1871 book *Theory of Heat*. It is there that he supposes that there's a container filled with air that possesses uniform pressure and temperature (the system is in equilibrium). The container is divided into two sides. The two sides of the container are labeled A and B, and the division between them is wrought by a diaphragm with a large hole in it. Over the hole is a sliding plate with very small mass that is controlled by "a being whose senses are so acute that he can see every molecule of the air, at least when it is near the hole".[94] Maxwell says the being follows the command: Open the plate over the hole when a molecule possessing "more than the mean velocity" in compartment A moves near the hole (n. 94). This allows for faster molecules to move into compartment B. The plate is to remain closed for all other molecules, although when in compartment B, a slower (*i.e.*, slower than the mean velocity) molecule draws near the opening, the plate is to be opened allowing that molecule to pass from B to A. Maxwell infers that compartment B will begin to enjoy an increase in the mean velocities of its inhabitant molecules, while compartment A will enjoy a decrease of mean velocities. These changes all obtain without the expenditure of work. As a result, "the [non-statistical and exceptionless] second law of thermodynamics is no longer true",[95] and "[t]he 2nd law of Thermodynamics has the same degree of truth as the statement that if you throw a tumblerful of water into the sea you cannot get the same water out again."[96]

Great. There were reasons to abandon a non-statistical and exceptionless statement of the second law before Boltzmann's proof of the H-theorem. But why should one think that Boltzmann was aware of those reasons when he tried to prove the H-theorem? Loschmidt articulated a Maxwell-"demon"-like case (without a demon) in 1869.[97] Boltzmann would have known of Loschmidt's version of the case since they were colleagues (*q.v.*, n. 97). Indeed, Boltzmann responds to Loschmidt, providing a version of the case that resembles Maxwell's. He wrote,

> When for instance a gas at constant temperature is divided into two halves by a separating wall with a small hole on it, it would be possible to bring in front of the hole a contraption that guides the faster molecules preferably into one half and the slower ones preferably into the other half, *which would contradict the second law*.[98]

#### 4.2.2 The Reversibility Objection before Loschmidt

I've argued that Boltzmann showed an awareness of the real possibility of the existence of gas systems to which the H-theorem could not be applied. My evidence for this resides beside the articulation of Boltzmann's original proofs of the H-theorem in his 1875 work. My second reason for maintaining that Boltzmann knew of failures of H-theorem application to even monatomic gas

---

94 (Maxwell, vol. 2: part 2 2009, 585).

95 (ibid.).

96 (ibid., 583). This is from his December 6th, 1870 letter to John William Strutt (Lord Rayleigh; 1842-1919). He affirms the quoted conclusion after presenting the "demon" case. In the version articulated in that letter, Maxwell says "I do not see why even intelligence might not be dispensed with and the thing [the sliding plate covering the hole] be made self-acting." (ibid.) See also the April 13th, 1868 letter to Mark Pattison, specifically at (ibid., 366-367) and (Maxwell 1902, 153-154).

For modern studies of Maxwell's "demon" case, see (Brush 1974, 40-41); (Darrigol 2018, 63-64); (Daub 1970) who notes the historical fact I'm noting here; (Klein, Demon 1970); (Smith 1998, 239-267); (Smith and Wise 1989, 621-626).

97 (Loschmidt 1869). See (Daub 1970, 218-221); (Garber, Brush, Everitt 1995, 57).

98 (Boltzmann 1870, 470), as quoted and translated by (Darrigol 2018, 182) emphasis mine.

systems had to do with his awareness of a Maxwell "demon" case prior to 1872. I now add that the reversibility objection (*q.v.*, **sect. 7** below) was articulated and probably known to Boltzmann before his proof of the H-theorem. This fact constitutes my third justification for believing that Boltzmann knew of H-theorem inapplicable systems before at least 1875. Notice that all three justifications support the further claim that Boltzmann was aware of nomologically possible violations of the non-statistical version of the second law prior to the 1875 publication of the H-theorem.

The already referenced 1867 letter from Maxwell to Tait included a penciled annotation by Thomson which read, "Very good. Another way [to violate the non-statistical second law of thermodynamics] is to reverse the motion of every particle of the Universe and to preside over the unstable motion thus produced."[99] The next year Maxwell said there was an apparent conflict between the existence of irreversible processes and the reversibility of all motion.[100] Two years later, Maxwell would write to Strutt arguing that reversing all of the motions (*i.e.*, velocities) of every particle in the universe would "upset the 2nd law of Thermodynamics."[101]

Of course, Boltzmann was quite far removed from Scotland and probably did not have access to the correspondence of Maxwell, Tait, or Thomson. Moreover, I can find no evidence that he was aware of the reversibility objection before the publication of (Boltzmann 1872). However, there is evidence that Boltzmann knew the work of Thomson, taking it seriously enough to cite it. For example, Boltzmann cited Thomson's work with Tait on the principle of least action in hydrodynamics at (Boltzmann, "On the Compressive Forces" 1871). In (Boltzmann 1874), Boltzmann cites Thomson once again, although this time on work related to electricity. We also know that much later, Boltzmann corresponded with Thomson in 1892 and 1893, answering several of Thomson's objections to Boltzmann's kinetic theory from energy dissipation.[102] I therefore think it is likely that Boltzmann read Thomson's reversibility objection in *Nature*, published in 1874. We know Boltzmann read *Nature* because Boltzmann published in it several times.[103] Why is this important? Because (again) in 1875, Boltzmann tried to prove his H-theorem in much the same way he did in 1872. However, this time, he'd seek to make his earlier work known to a broader audience (Darrigol 2018, 172). His remarks regarding the non-statistical nature of the second law *as viewed through the lens of* the H-theorem are univocal:

> [We] have so far proceeded as follows: [we] have shown that the quantity H cannot increase during the evolution of the state of the gas; wherefrom [we] have concluded that it must be constant in the case of equilibrium since it evidently cannot constantly decrease in this case. I was thus able to derive the definitive equations that lead to the equilibrium distribution of states. This suggests that the value that H takes in the case of equilibrium is the smallest of all the values that H can take in agreement with the conservation of the total number of atoms and the conservation of the total live force.[104]

---

[99] (Knott 1911, 214); *cf.* (Smith and Wise 1989, 625). Thomson would follow-up on this thought in (Thomson 1874).

[100] (Maxwell, vol. 2. Part 1 2009, 361).

[101] (Maxwell, vol. 2: Part 2 2009, 582).

[102] See the correspondence cited in (Smith and Wise 1989, 428-429 n. 81).

[103] (Boltzmann, Certain Questions 1895); (Boltzmann, [Reply to Culverwell] 1895); (Boltzmann, Minimum Theorem 1895).

[104] (Boltzmann, "On the Thermal Equilibrium of Gases on Which External Forces Act" 1875); *BWA*2, 22-23. Taken from the translation work of (Darrigol 2018, 175).

If Boltzmann was aware of the reversibility objection before 1875, why would he assert the above if reversibility worries should cause him to abandon deterministic or non-statistical statements of minus-H increase over time? I believe the best charitable response to this query should not argue that Boltzmann had a statistical interpretation *of the H-theorem* all along (*contra* (Badino 2011), *q.v.*, my arguments against this in n. 81 and in **sect. 5.3** below). Rather, the correct response suggests instead that Boltzmann believed that while the deterministic or non-statistical statement of the second law of thermodynamics admits exceptions, at least some of those exceptions (if not all of them) have to do with systems to which the H-theorem cannot be applied. However, if the H-theorem does apply to a system out of equilibrium, minus-H *must* increase monotonically for all time until equilibrium is obtained. There it will remain *given that the Boltzmann equation holds for all time* (anticipating here worries about fluctuations and recurrence about which I will say more in a part two essay). You see, proof of the H-theorem amounts to proof of a deterministic and non-statistical second law *suitably restricted* to systems that satisfy the antecedent of the H-theorem. Of course, my reading assumes that Boltzmann strongly associates H (or minus-H) with entropy. Let me now say more about that association in Boltzmann's corpus.

### 4.3 Interpreting the H-Theorem: $-$H and Entropy

Boltzmann's 1872 interpretation of the H-theorem is given in the following often-quoted passage which I will call **PERICOPE**,

> It has thus been rigorously proved that, whatever may be the initial distribution of kinetic energy, in the course of a very long time it must always necessarily approach the one found by Maxwell [notice that he's speaking here in terms of energy changes]. The procedure used so far is of course nothing more than a mathematical artifice employed in order to give a rigorous proof of a theorem whose exact proof has not previously been found. It gains meaning by its applicability to the theory of polyatomic gas molecules. There one can again prove that a certain quantity [H] can only decrease as a consequence of molecular motion, or in a limiting case can remain constant. One can also prove that for the atomic motion of a system of arbitrarily many material points there always exists a certain quantity which, in consequence of any atomic motion, cannot increase, and this quantity agrees up to a constant factor with the value found for the well-known integral $\int \frac{dQ}{T} = 0$ in my paper on the 'Analytical proof of the 2nd law, etc.'. We have therefore prepared the way for an analytical proof of the second law in a completely different way from those previously investigated. Up to now the object has been to show that $\int \frac{dQ}{T} = 0$ for reversible cyclic processes, but it has not been proved analytically that this quantity is always negative for irreversible processes, which are the only ones that occur in nature. The reversible cyclic process is only an ideal, which one can more or less closely approach but never completely attain. Here, however, we have succeeded in showing that $\int \frac{dQ}{T}$ is in general negative, and is equal to zero only for the limiting case, which is of course the reversible cyclic process (since if one can go through the process in either direction, $\int \frac{dQ}{T}$ cannot be negative).[105,106]

**PERICOPE** suggests that minus-H is entropy, and that H is equal to minus entropy.[107] It implies that the H-theorem is Boltzmann's attempt to ground the second law of thermodynamics in mechanics or microdynamics.[108] But why does Boltzmann use minus-H and $\int \frac{\delta Q}{T}$ (which I will call the *Clausius integral*) interchangeably? The question is perplexing because the entropy of Boltzmann's 1872 and 1875 work is that of a *closed system*, while the Clausius integral has to do with heat *Q* being exchanged with a system at absolute temperature *T*. To properly answer this question, we must first understand Rudolf Clausius's theory of entropy.

# 5 Clausius Entropy

## 5.1 The Clausius Integral and Entropy

Clausius associated the entropy of physical systems, *at least in certain contexts*, with the *Clausius integral* in his 1865 memoir[109], where the differential $\delta Q$ is a differential form representative of exchanged heat or infinitesimal heat transfers into the system as part of a reversible process. The source of the heat is some external system enjoying absolute temperature *T*. For reversible cyclic evolutions, we have:

(10):

$$\oint \frac{\delta Q}{T} = 0$$

$\frac{\delta Q}{T}$ is therefore an exact differential. Clausius likewise tried to capture what he understood to be irreversible processes involving systems that evolve from one equilibrium state to another equilibrium state in a manner that could be reversed by means of a reversible transformation. While the relevant transformations in the cycle obtain, the inequality given by (11) holds true:

(11):

$$\int \frac{\delta Q}{T} < 0$$

The total entropy of the *global* system that includes external sources for the needed heat exchange was said to increase in entropy (hence, an *irreversible* process). The entropic increase on account of a *reversible* transition of a system from one equilibrium state $S_1$ to a distinct equilibrium state $S_2$ is less than the entropy increase on account of an *irreversible* transition from one equilibrium state $S_3$ to a distinct equilibrium state $S_4$.[110] In (Clausius 1866, 5), Clausius would add (12) below understanding the conjunction of it with (10) as an expression of the two fundamental equations of thermodynamics.

---

[105] (Boltzmann 1966, 117). I will use '$\delta Q$' to mean exchanged heat. It stands in for Boltzmann's use of '$dQ$'. Below I will also use the expression '$\delta W$' to mean exchanged work.

[106] See the discussion of some of the ideas here in (Cardwell 1971, 266-267), and (Segrè 1984, 231).

[107] Well, for an ideal gas, entropy equals minus *kHV* (Davies 1977, 42).

[108] As Segrè noted, "[u]ltimately, Boltzmann showed that H was the negative of the entropy. He had thus connected thermodynamics with mechanics, but through the roundabout way of the H-theorem." (Segrè 1984, 243)

[109] (Clausius 1865). The integral actually first appears in (Clausius 1854). *Q.v.*, n. 116. My reading of Clausius follows (Cardwell 1971); (Cercignani 1998, 80-85); (Darrigol 2018, 42-50); and (Daub 1967, 301-303) in some places, but it also depends heavily upon my own independent assessment of the work of Clausius.

(12):

$$\delta Q = dU + \delta W$$

where $U$ is energy[111] and $\delta W$ represents the infinitesimal process of work done while heat is transferred.[112] $Q$ and $W$ are here path dependent quantities, while $U$, like entropy $S$, is a path independent state function or property. Clausius would also add that,

(13):

$$\frac{\delta Q}{T} = \frac{dH}{T} + dZ$$

where $H$ and $Z$ are both state functions (the latter by stipulation really), $H$ being the heat of the body/system although he would famously reduce that notion to *vis viva* (kinetic energy) thereby reducing it to motion[113], and $Z$ being disgregation. More on these two quantities soon. Reflect, for now, on the fact that because (13) holds, and because Clausius was strongly associating entropy with $\frac{\delta Q}{T}$, in 1865, Clausius affirmed that:

(14):

$$dS = \frac{dH}{T} + dZ$$

And because (13) holds, we have shown precisely *how* Clausius strongly associates entropy with heat exchange. Indeed, one popular way of charactering Clausius's understanding of the second law is as follows:

> Heat cannot pass spontaneously from a body of lower temperature to a body of higher temperature.[114]

[110] (C. Cercignani 1998, 82). See also (Darrigol 2018, 47-48); (Müller and Weiss 2005, 1-12).

[111] In *contemporary* discussions, $U$ represents internal energy. For the idea in Clausius, see (Clausius 1854); (Clausius 1866); (Clausius 1879); (Daub 1967, 293-294); and the ensuing discussion in the main text below.

[112] Even in contemporary physics, equation (12) is also commonly understood to be a mathematical expression of the first law of thermodynamics (Alonso and Ydstie 1996, S1119); (Wood 1981, 311).

Clausius remarked, "work may transform itself into heat, and heat conversely into work, the quantity of one bearing always a fixed proportion to the other." (Clausius 1879, 23 emphasis removed). For idealized gases and fluids, $dW = pdV$. This equation expresses the equivalence of heat and work, the very principle discovered by Julius Robert Mayer (1814-1878) and James Prescott Joule (1818-1889).

[113] "…a motion of the particles does exist, and that heat is the measure of their *vis viva*." (Clausius 1851, 4).

[114] (Klein and Nellis 2012, 207 emphasis removed).

Clausius himself remarked under a section entitled "New Fundamental Principle concerning Heat", "Heat cannot, of itself, pass from a colder to a hotter body,"[115] and as early as 1854, Clausius declared that he had found an analytic "expression of the second law" for reversible cyclic processes in equation (10).[116] These equations and remarks from Clausius flirt with Maxwell's view of entropy and heat exchange in his *Theory of Heat*. Maxwell said, "when there is no communication of heat this quantity [entropy] remains constant, but when heat enters or leaves the body the quantity increases or diminishes."[117] I believe it is a mistake to attribute Maxwell's position to Clausius—Clausius's strong association of entropy with the relevant integral notwithstanding—for two reasons. First, in 1865, Clausius's statement of the second law referred to the entropy of the universe. The universe, however, is a closed isolated system that does not exchange heat with some other body. Moreover, he says that the entropy of the cosmos increases to a maximum. Not all subsystems of an entropy increasing cosmos such as ours will be in equilibrium. Thus, one cannot explain the increasing entropy of our cosmos by appeal to equilibrium subsystems exchanging heat with their environments.

Second, Maxwell's statement was experimentally falsified by the Gay-Lussac—Joule experiment,[118] and there are good reasons to believe Clausius was aware of the relevant experimentation, for Clausius was familiar with the work of Joseph Louis Gay-Lussac (1778-1850), having discussed his work several times in *The Mechanical Theory of Heat*. He was also familiar with Joule's experimentation, citing his results authoritatively throughout the same work.

### 5.2 The Complete Nature of Clausius Entropy

What then *is* the complete and fully general nature of entropy according to Clausius? He clearly had some more general conception, for *again*, in one place he characterizes the second law of thermodynamics as the principle that "[t]he entropy *of the universe* tends to a maximum"[119], christening it (*i.e.*, the law itself) with fundamental status.[120]

Clausius's seminal work on thermodynamics was his 1864 *Abhandlungen über die mechanische Wärmetheorie* (*Treatises on the Mechanical Theory of Heat*) which was later (1876) to become a more "connected whole" under the title *The Mechanical Theory of Heat* (I have

---

[115] (Clausius 1879, 78). He allowed for the passage of heat from cold to warmer bodies so long as there was a simultaneously occurring compensating process.

[116] (Clausius 1854, 500),

> "Demnach gilt für alle *umkehrbaren Kreisprocesse* als analytischer Ausdruck des zweiten Hauptsatzes der mechanischen Wärmetheorie die Gleichung (II.) $\int \frac{\delta Q}{T} = 0$." (ibid. emphasis in the original) worked and will work with the 1879 English translation).[121] In that work, the entropy of a thermodynamic system is not as fundamental as that system's energy $U$ *suitably understood* (Clausius 1879, 195). I include the qualification because in the 1867 (first English edition) of *The Mechanical Theory of Heat* (Clausius 1867), Clausius agreed with William Rankine (1820-1872), maintaining that small changes of energy are given by the sum of small changes to $H$ and small changes to internal work ($I$) due to internal molecular forces, and so, $dU = dH + dI$.[122] In the improved and transmuted 1879 edition, $U$ is "[t]he sum of the Vis Viva and of the Ergal…"[123], but Ergal ($J$ in Clausius's work) is just potential energy (first coined by Rankine).[124] The type of energy involved here appears to be both kinetic and potential energy. And this notion of energy ($U$) is one and the same as that in the work of Thomson. Clausius related his conception to Thomson's as early as 1866 (Clausius 1866, 5).

[117] (Maxwell 1902, 162). The remark is repeated twice at (ibid., 190 and 191).

[118] See (Cardwell 1971, 276); (Cheng 2006, 142-144). The experiment involved a large container with two chambers separated by a diaphragm. The container features only thermally insulated walls cutting off all heat exchange between the container's contents and the container's environment. A gas is introduced into one of the chambers, and the diaphragm subsequently released. A free expansion takes place. As the gas expands, evolving adiabatically (no heat exchange!), no work is performed, no temperature change takes place, *and yet entropy increases*.

[119] From (Clausius, Concerning Various Convenient Forms of the Main Equations of the Mechancial Theory of Heat 1865), as translated by (Cardwell 1971, 273 my emphasis).

[120] He calls both the first and second laws of thermodynamics "fundamental laws of the universe" and "fundamental theorems of the mechanical theory of heat" in his 1865 memoir. See (Cardwell 1971, 272-273) and the citations and quotations therein.

More should be said because in the 1867 edition of *The Mechanical Theory of Heat*, and the original work behind it, there exists a more fundamental quantity lurking beneath entropy, *viz.*, disgregation. Disgregation ($Z$ in Clausius's equations) is a quantity increased by heat. In fact, for Clausius, heat causally produces work as in (Carnot 1960) and it does this by increasing disgregation.[125] Disgregation itself is "the degree in which the molecules of a body are dispersed".[126] In this earlier edition of Clausius's major work, entropy, or infinitesimal changes thereof, was/were specified by appeal to, *inter alia*, disgregation or infinitesimal changes thereof. We already expressed the mathematics that encodes these facts via equation (14). As I've already said, the quantity $H$ in that equation is the heat in the system, but by this Clausius meant *vis viva* or kinetic energy of molecular motion, as Maxwell (Maxwell 1878, 258) pointed out criticizing Tait's misreading of Clausius.[127] This treatment (*i.e.*, (14)) of entropy was lambasted by both Maxwell and Tait, the former providing the clearer and more measured response of the two.[128] As I already foreshadowed, Clausius removes disgregation from the later 1879 edition of *The Mechanical Theory of Heat*. But even if we keep $Z$ in Clausius's framework, the principle cause of the increase of disgregation is molecular motion responsible for increasing dispersion, and that motion can be understood in terms of kinetic energy. Thus, $U$ (which for Clausius you'll recall is Vis Viva and Ergal) resides beneath disgregation (*i.e.*, it is more fundamental than disgregation) which is related to entropy in the way (14) suggests.

Having discovered Clausius's mature view on the status of entropy, energy, and disgregation, in the hierarchy of being, we should now answer the question: What, according to Clausius, is entropy? Entropy is that property of physical systems that tracks (or serves as a measure of) the processes of energy transformation, often, though not always associated with heat exchange (Cardwell 1971, 273). He wrote,

[121] (Clausius 1879).
[122] (Daub 1967, 293-294).
[123] (Clausius 1879, 20).
[124] (Clausius 1879, 11), *cf.* (Daub 1967, 293-294); (Daub 1969).
[125] "…the increase of disgregation is the action by means of which heat performs work…" (Clausius, Application of the Theorem 1862, 91).
[126] (Clausius 1867, 220). *Cf.* (M. J. Klein 1969, 136). The notion of disgregation has a not too distant cousin in prior work by Rankine. The relevant notion in Rankine's work is the metamorphic function.
[127] (Daub 1967, 293-294).
[128] (Maxwell, Tait's "Thermodynamics" 1878). Boltzmann thinks he has proven the existence of Clausius's disgregation in (Boltzmann, Analytical Proof of the Second Principle 1871); (Darrigol 2018, 130).

> We might call '*S*' the transformational content of the body, just as we termed the magnitude '*U*' the thermal and ergonal content. But as I hold it better to borrow terms for important magnitudes from the ancient languages so that they may be adopted unchanged in all modern languages, I propose to call the magnitude *S*, the entropy of the body, from the Greek word τροπη, transformation.[129]

For Clausius, there "is a natural bias in the distribution of energy and in the direction which energy changes tend to take. Entropy gives us a measure of this bias in the case of material bodies or systems of bodies"[130]

### 5.3 Boltzmann and the Clausius Entropy

There can be no doubt that Boltzmann knew Clausius's work on entropy, and that his understanding of entropy was, *in some places*, one and the same as that of Clausius. For example, Boltzmann's second published paper was "On the Mechanical Significance [Meaning] of the Second Law of Heat Theory".[131] There, perhaps under the influence of (Rankine 1865), Boltzmann tries to provide an extension of Clausius's earlier conception of entropy to systems that feature molecules that enjoy periodic motions assuming all the while many aspects of Clausius's early kinetic theory of gases. That it is Clausius's concept of entropy that Boltzmann is extending is well justified by the fact that in 1871, Clausius would provide the same extension of *his* (*i.e.*, Clausius's) concept to periodic motions (Clausius, Reduction of the Second Law 1871), realizing *after Boltzmann's rebuke*, that Boltzmann's extension or generalization of Clausius's concept came before Clausius's own generalization in (Clausius, Remarks on the Priority Claim of Mr. Boltzmann 1871).[132] Furthermore, $dS = \frac{\delta Q}{T}$ holds in both of the aforementioned 1871 papers by Boltzmann and Clausius. Of course, this is the case in Clausius's earlier work too.[133]

Later review and correspondence included at least two concessions by Boltzmann. Clausius in (Clausius, Reduction of the Second Law 1871) accomplished something that Boltzmann in (Boltzmann 1866) failed to. Clausius provided the more accurate mathematical characterization of entropy, and (Boltzmann 1866) had ignored changes in the potential.[134] These concessions suggest that Boltzmann took Clausius's work on entropy quite seriously. But one might now ask:

> (**Maxwell's Question**): Why would Boltzmann worry about priority in this context [Boltzmann rebuked Clausius for reproducing Boltzmann's earlier work] when Boltzmann had already begun exploring the Maxwellian distribution-based approach to thermodynamics and statistical mechanics?

---

[129] As quoted and translated by (Cardwell 1971, 272). Taken from (Clausius 1865, 353). This understanding does not go away in his later work. See, *e.g*., (Clausius 1879, 107).

[130] (Cardwell 1971, 272).

[131] (Boltzmann 1866).

[132] (C. Cercignani 1998, 83-84); (Darrigol 2018, 108-109). That Boltzmann is extending Clausius's concept of entropy is the opinion of Olivier Darrigol (Darrigol 2018, 70), though he uses the phrase "counterpart to Clausius's entropy in periodic mechanical systems".

[133] The idea behind both of Boltzmann and Clausius's 1871 papers was to extend this relation to systems featuring periodic molecular motion, and so additional terms are expressed via additional equalities. See (Darrigol 2018, 108).

[134] See (Darrigol 2018, 109) for the details. Darrigol adds, "I agree with Clausius that Boltzmann's derivation of the second equation implicitly excludes a change in the potential function." (ibid.)

I call this **Maxwell's question** because in an 1873 letter to Tait, Maxwell would ponder a similar query, expressing his astonishment about the continued interest in mechanical approaches to the second law. He wrote,

> It is a rare sport to see those learned Germans contending for the priority of the discovery that the 2nd law of [thermodynamics] is the [Hamilton Principle]… the [Hamilton Principle], the while, soars along in a region unvexed by statistical considerations, while the German Icari flap their waxen wings in nephelococcygia [*i.e.*, cuckoo land] amid those cloudy forms which the ignorance and finitude of human science have invested with the incommunicable attributes of the Queen of Heaven.[135]

The answer to Maxwell's question is simple. Boltzmann cares about mechanical justifications of the second law and regards his H-theorem as a means whereby he can attain just such a mechanical justification. That is why in **PERICOPE** he speaks, albeit rather imprecisely and somewhat clumsily, of entropy as if it is represented by the Clausius integral. What Boltzmann is actually attempting to do is connect minus-H with Clausius's energy transformational notion of entropy, the type of entropy which Clausius believed had a mechanical explanation. *Boltzmann's H-theorem should be viewed as the fulfillment of Clausius's vision*. The H-theorem does in fact provide a mechanical explanation of how energy tends to transform and, more derivatively, how entropy tends toward a maximum.[136]

That Boltzmann is borrowing Clausius's transformational conception of entropy in his work on the H-theorem is a conclusion other scholars have reached. Olivier Darrigol and Jürgen Renn state,

> Boltzmann…noted that the value of −H corresponding to Maxwell's distribution was identical to Clausius's entropy. For other distributions, he proposed to regard this

---

[135] As quoted by (Darrigol 2018, 109). Boltzmann claimed in (Boltzmann, On the Mechanical Significance [Meaning] of the Second Law of Heat Theory 1866) to have derived a Clausiustical entropy law from the principle of least action, hence Maxwell's reference to Hamilton's principle.

[136] As further evidence for the claims in the main text, consider the fact that the paper Boltzmann references in **PERIOCOPE** above is (Boltzmann, Analyatical Proof of the Second Principle 1871). There Boltzmann attempted to specify the entropy of a system that exemplifies what's called the canonical distribution or $\rho(x) = e^{-\beta H(x)} / \int e^{-\beta H} d\sigma$ (where $d\sigma$ provides the phase orbit invariant measure on the phase space used to model the system, and where the H(x) here is the phase (all positions and velocities of the atoms in the system) dependent energy of the same system). Here Boltzmann is reaching back to his earlier attempt in (Boltzmann, On the Mechanical Significance [Meaning] of the Second Law of Heat Theory 1866) to provide a mechanical explanation of the second law of thermodynamics. He thought that he could apply the notion of Clausius entropy to gas systems featuring molecules that enjoy periodic motions, subsequently coming to understand that he won't be able to explain non-periodic gas systems. He follows (and this is further evidence that he's working with Clausius's understanding of entropy) (Clausius, Remarks on the Priority Claim of Mr. Boltzmann 1871) in his attempt to derive Clausius's notion of disgregation. He uses that bit of ideology with its underlying concept to retrieve the accepted minus integral expression for the entropy of a system abiding by the canonical distribution, noting along the way, that one could relate or associate entropy with kinetic and potential energy. Like Clausius, Boltzmann is here plainly describing changes in entropy in terms of transformations of energy, and he is fully embracing not only Clausius's notion of entropy but also Clausius's notion of disgregation! See (Darrigol 2018, 128-133, on which I lean) for additional commentary.

> function as an extension of the entropy concept to states out of equilibrium, since it was an ever-increasing function of time.[137]

But I go further than Darrigol and Renn because there is an additional inference to make. Because the H-theorem has to do with Clausius entropy, and Clausius entropy is defined in terms of energy, tracking how energy transforms over time, we can agree with noted historian Stephen G. Brush when he says that "the H-theorem is a microscopic version of the general principle of dissipation of energy proposed by Kelvin in 1852, and reformulated by Clausius in 1865 in the phrase, 'the entropy of the universe tends to a maximum.'"[138] In other words, we can affirm that according to both Clausius *and Boltzmann's* deep conception of entropy, entropy is that quantity which tracks the way energy is changing or transforming over time. This deep conception of entropy understood as a quantity that tracks energy can be used to describe thermodynamic systems both in and out of equilibrium. The H-theorem not only helps facilitate such descriptions it also helps provide a mechanistic explanation of entropic increase and stability after equilibrium is reached.

Understanding Boltzmann's H-theorem as a theorem about a type of entropy that tracks transformations of energy is not new. Glimmers of that interpretation of Boltzmann appear in the work of Edward P. Culverwell's (1855-1931) 1890 article on Boltzmann's kinetic theory. Culverwell explicates Boltzmann's characterization of a gas in equilibrium as a system enjoying a status which entails that that gas's "energy is equally distributed among all…[its] degrees of freedom".[139] Moreover, this precise entropy-energy connection is used in the contemporary practice of thermodynamics. As Klein and Nellis put it in their recent textbook on thermodynamics, "the property entropy is introduced in order to quantify the quality of energy."[140]

## 6 The Probabilistic Interpretation of the H-Theorem

Given the reasoning of **sect. 5**, it may be difficult to see room for the probability calculus and its accompanying interpretation since the explanation the H-theorem affords is mechanical. According to **sect. 4.1**, for Boltzmann (as for Clausius and Maxwell) a mechanical explanation is one that explains features of a gas system SYS by appeal to the causal behavior of subsystems of SYS. As in Maxwell's statistical mechanics, probability does enter Boltzmann's reasoning. It does so in a way that manifests an epistemic view of probability. According to Boltzmann, we don't know the precise state of the molecules of SYS, so we cook up our best understanding of how they are dancing (a statistical distribution law) and then, given that assumption, we look to mechanical features to see how the subsystem's velocities are changing as they approach the state described by the distribution law. The approach to that state, as well as the mechanism whereby the system remains in that state has directly to do with causal influences among the subsystems of SYS. Appreciating the velocity changes due to causal influences revealed in the collisions that push a non-equilibrium system like SYS toward equilibrium is the means whereby we appreciate how the system's entropy is changing over time. Our entire methodology is always removed from the precise details about actual world goings-on because the way we are modeling the involved causal influence through velocity change is through equations about how the distribution function itself changes over time. Our best efforts can only ever be approximate, and time has told us (or, from Boltzmann's perspective, will tell us) that statistical hypotheses coupled with the right equations (*i.e.*, the Boltzmann equation and a statement of the H-functional) bear fruit and aid us in our efforts to save the phenomena.

[137] (Darrigol and Renn 2013, 773). See also (Jungnickel and McCormmach 1986, 64); (Kuhn 1978, 42); (M.J. Klein, Development 1973, 68).
[138] (Brush vol.1 1976, 80).
[139] (Culverwell 1890, 96). More contemporaries of Boltzmann could be cited.
[140] (Klein and Nellis 2012, 237).

Obviously, the aforementioned statistical hypotheses include quantitative statements of equations revealing the contents of distribution functions like the Maxwell and Maxwell-Boltzmann distribution laws. That these laws were understood by Boltzmann to give probabilities is evidenced by both Boltzmann's interpretation and *reinterpretation* of the Maxwell distribution in (Boltzmann, "Studies on the Equilibrium of Live Force Between Moving Material Points" 1868)[141], and in Boltzmann's first attempt at proving the H-theorem in 1872. Beside his 1872 attempted proof is an agreement with Maxwell. Probabilistic methods are required (quoting Darrigol's careful exegesis) "in order to deal with highly irregular processes involving a huge number of molecules. The irregularity and the law of large numbers explain the stability of macroscopic averages."[142]

Does the admission of probabilistic resources into the H-theorem project mean that there are exceptions to the theorem? That is a tricky question. Theorems are necessarily true, if true. Given that the antecedent of a theorem is satisfied, the consequent is *strictly* implied. There are no exceptions to the H-theorem in one important sense then. But while Boltzmann does show, for systems satisfying the antecedent of the theorem, that the distribution of a gas system SYS at a time $t$ causally depends (the Boltzmann equation is a deterministic equation) upon its distribution at some prior time, (as I've said before) the notion of a distribution itself is statistical or probabilistic.[143] There is no guarantee that nature always gives us systems fit for the assumption that the distribution used is appropriate (*q.v.*, my discussion along these lines earlier in **sect. 4.2**), and it is for this reason that I (and more importantly Boltzmann) have already said that there can be systems that do not evolve in the way demanded by the Boltzmann equation and the H-theorem.

### 6.1 Loschmidt's Reversibility Objection: Articulated

Perhaps there is a way of more direclty objecting to the use of the H-theorem in attempts to mechanically explain appearances. Recall that the antecedent of the H-theorem is the conjunction that identity (8) holds, the Boltzmann equation is omnitemporally true, and the distribution function $f$ satisfies that equation.[144] We can now appreciate the question: what if we could find a system that satisfied the antecedent of the theorem but which did not have a distribution that tended toward the Maxwell distribution? An objection along these lines was voiced by Loschmidt (J. Loschmidt 1876). His objection (the reversibility objection) began with the correct assumption that the laws of classical mechanics (specifically Hamiltonian mechanics) are time-reversal invariant and that therefore the evolutions involving the increase of the minus-H function can be turned around resulting in evolutions involving a decreasing minus-H function. These minus-H function decreasing evolutions are perfectly consistent with the underlying laws of Hamiltonian mechanics because those laws are time-reversal invariant. If, however, the mechanics drives the minus-H function increase, then how can minus-H decrease be driven by the

141 See (Darrigol 2018, 529); (Uffink 2007, sect. 4.1).

142 (Darrigol 2018, 531). But see *BWA*1, 316-317 for these ideas in Boltzmann.

143 Agreeing with this nice point in (Darrigol 2018, 532).

144 As will become clear, the theorem must also make use of the **HMC** (*q.v.*, **sect. 7** for a definition).

same mechanics? The reversed evolutions contradict the H-theorem. This is the reversibility paradox.[145]

### 6.1.1 The H-Theorem Untarnished in Boltzmann's Eyes

My response to Loschmidt's famous objection will come later (**sect. 7**). For now, I point out that contrary to what the Standard Story would have us believe, after Boltzmann engaged with Loschmidt's reversibility objection, he continued to positively affirm the H-theorem as a means whereby one achieves insight into the deterministic and exceptionless increase of entropy *for systems that satisfy the antecedent of the H-theorem*. He continued on in this way late into his career, viewing his H-theorem as the mechanical justification of the second law, even after articulating his combinatorial definition of entropy and his combinatorial arguments for a statistical statement of the second law. The H-theorem was viewed by him to be a more fundamental justification of the second law, one which the combinatorial arguments illustrate.[146] There are four reasons in favor of this understanding of Boltzmann's work.

First, in Boltzmann's very reply to Loschmidt, Boltzmann affirms that "the existence of microstates for which the entropy decreases does not contradict the general endeavor to *deduce* the entropy law from atomistic considerations" (Darrigol 2018, 198 my emphasis). The very section immediately following Boltzmann's reply to Loschmidt is entitled "Comments on the *Mechanical* Meaning of the Second Law of Heat Theory". There Boltzmann invests time and energy discussing the mechanical justification of the second law, never giving up on it.

Second, as late as Boltzmann's first volume of the *Lectures on Gas Theory* (1896), Boltzmann says, "if at the beginning of some time interval [the value of the distribution function] is on the average the same at each position in the gas…, the same will hold true at all future times".[147] And he would add, "the quantity we have called H can only decrease."[148] In the second volume of the *Lectures on Gas Theory* published in 1898, after churning through several points in a proof sketch, Boltzmann concluded, "[s]ince the same holds for all other kinds of molecules, and similarly for collisions of different molecules of the same kind with each other, we have proved

---

[145] See (Loschmidt 1876); (Boltzmann, Comment on Some Problems in Mechanical Heat Theory 1877); (Darrigol 2018, 198).

[146] As Olivier Darrigol said in correspondence,

> "If you (and Kuhn) mean that for Boltzmann the combinatorial entropy formula was not primitive and that the Boltzmann equation and the equilibrium theorems were in the end more important, I completely agree. For a couple of years after 1877 he seems to have believed that he had a new way to compute thermodynamical equilibrium with this formula. But he later realized (in 1881) that the formula [was] in fact derived from the better founded microcanonical distribution. *In the lectures on gas theory, the combinatorial entropy formula is there only as a 'mathematical illustration' of the H function, which is introduced through the Boltzmann equation and the H theorem*." (11/19/2019 emphasis mine)

Badino raises an important question that few in the literature have sought to answer, "…if it is true that Boltzmann in 1877 abandoned a strict mechanistic view in favor of a probabilistic one, why did he consistently keep using the 1872 approach in his publications throughout the rest of his life?" (2011, 354-355). I believe that Badino and I would reply that he never abandoned the mechanistic view but our attitudes about how best to understand Boltzmann's views about mechanics, probability, the H-theorem, and the combinatorial arguments, differ substantially.

[147] As translated and quoted by (Kuhn 1978, 45).

[148] (Boltzmann 1964, 55).

that in this special case the value of H can only decrease as a result of collisions."[149] Undoubtedly, throughout Boltzmann's corpus, the way he views the H-theorem and its implications is "predominantly deterministic".[150]

Third, when Boltzmann was writing his *Lectures on Gas Theory*, he stopped midway through. Why did he do this (volume 1 was published in 1896, volume 2 in 1898)? He did it because he thought it necessary to author a treatise on mechanics *because his gas theory was/or should be grounded (he believed) in mechanics*. So, he published volume one of his treatise on mechanics in 1897. As he says in his *Lectures on Gas Theory*, the atomistic approach to the physics of matter provides the best ***mechanical*** explanation of nature (1964, 26-27). This was no isolated supposition in Boltzmann's more general corpus. As Jungnickel and McCormmach (1986, p. 191) state, "Boltzmann presented mechanics as the foundation of *all theoretical physics*."[151]

Fourth, Boltzmann's general physical methodology distinguished between mechanical principles or laws, hypotheses, and the world. The laws are those of classical Hamiltonian mechanics. Hypotheses are principles like the second law of thermodynamics. Laws or mechanical principles are tested by the confirmation or disconfirmation of the hypotheses *they entail*. Hypotheses are confirmed in conjunction with the mechanical laws from which they follow.

> …neither the Theory of Gases nor any other physical theory can be quite a congruent account of facts…Certainly, therefore, Hertz is right when he says: 'The rigour of science requires, that we distinguish well the undraped figure of nature itself from the gay-coloured vesture with which we clothe it at our pleasure.' But I think the predilection for nudity would be carried too far if we were to forego every hypothesis. Only we must not demand too much from hypotheses…[152]

He continued:

> *Every hypothesis must derive indubitable results from mechanically well-defined assumptions by mathematically correct methods*. If the results agree with a large series of facts, we must be content, even if the true nature of facts is not revealed in every respect. No one hypothesis has hitherto attained this last end, the Theory of Gases not excepted. But this theory agrees in so many respects with the facts, that we can hardly doubt that in gases certain entities, the number and size of which can roughly be determined, fly about pell-mell. Can it be seriously expected that they will behave exactly as aggregates of Newtonian centres of force, or as the rigid bodies of our Mechanics? And how awkward is the human mind in divining the nature of things, when forsaken by the analogy of what we see and touch directly?[153]

---

[149] (Boltzmann 1964, p. 421 emphasis mine; see also page 432 where he says, "Hence dH/dt will be negative, and can be zero only when the condition (266) is satisfied for all collisions.")

[150] (Kuhn 1978, 57).

[151] Jungnickel and McCormmach would go on to point out that Boltzmann seems to judge that the old mechanical picture was starting to be superseded by a "new atomistic picture" (1986, 191). But that picture is not provided by statistical mechanics. It is provided by "modern electron theory".

[152] (Boltzmann, Certain Questions 1895, 413).

[153] (Boltzmann, Certain Questions 1895, 413-414) emphasis mine. *Cf.* (Darrigol 2018, 373).

This is a hypothetico-deductive method that includes the humble assertion that nature may not conform perfectly to our hypotheses and mechanical laws. According to this method, hypotheses like the second law *must follow* from mechanics.

While Boltzmann (Certain Questions 1895) provides me with some ammunition for my reading, the same source could be interpreted as completely taking it away:

> It can never be proved from the equations of motion alone, that the minimum function H must always decrease. It can only be deduced from the laws of probability, that if the initial state is not specially arranged for a certain purpose, but haphazard…the probability that H decreases is always greater than that it increases.[154]

This quotation gives my exegetical project the most serious kind of trouble. In it, Boltzmann admits to being unable to recover the H-theorem from mechanical considerations and suggests that the relationship between the antecedent of the *theorem* and its consequent is a probabilistic relation. I find that this series of remarks contains elements that are false, and worse, nonsensical. Again, the antecedents of theorems *entail* their consequents, and yet Boltzmann is quite clearly allowing for cases in which the consequent fails to follow from satisfaction of its antecedent. That is nonsensical. In addition, Boltzmann says that the H-theorem does not follow from mechanics. But I have already pointed out how Lanford showed that on the supposition that a choice gas system is dilute and that its constituents are approximated as hard shells (plus some further assumptions), the Boltzmann equation follows from the time-reversal invariant equations of motion in classical mechanics.[155] As Fields medal winner Cédric Villani stated at his 2010 Cambridge University lecture,

> Probably the single most important theorem in the [kinetic] theory remains the Lanford theorem from 1973. Lanford rigorously derived the Boltzmann equation from Newtonian mechanics…[for an appropriate domain]…in…[the appropriate limit] you recover the Boltzmann equation…This was the first result showing that you could…get this Boltzmann equation out of the Newton equation[s].[156]

Proofs of the H-theorem itself have been articulated in such a way that they satisfy the standards of rigor in contemporary mathematics (*q.v.*, n. 79).

We should not take the passage quoted above and cited in note 157 too seriously despite how often it is quoted. Boltzmann contradicts it numerous times in his *Lectures on Gas Theory*, and those lectures are the best source for Boltzmann's mature thought on thermodynamics and statistical mechanics. Here are the reasons for this:

---

[154] (Boltzmann, Certain Questions 1895, 414).

[155] Again see (Lanford 1975); (Lanford 1976); and (Lanford 1981). More precisely, what Lanford showed was that in the Boltzmann-Grad limit and for systems approximated by the hard sphere model, given smallness of time, that a particular weak chaos property holds initially, and some other assumptions, one can move from the BBGKY formulation or hierarchy (of equations) to the Boltzmann equation, itself formulated in terms of a hierarchy (the Boltzmann hierarchy). Of course, the BBGKY can be connected to Hamiltonian mechanics. For that, see (Uffink and Valente 2010, 147-150).

Again see (Lanford 1975); (Lanford 1976); and (Lanford 1981). But see also my comments on the relevant result in footnote 70.

[156] (Villani, Lecture 2010); (Villani, Lecture Notes 2010, slide 17); *cf*. (Villani, Math Berlin 2010).

(a) In 1900, the minister in Vienna described Boltzmann's attempt to acquire recognition and leadership status among the community of physicists through the publication of his *Lectures* (both his lectures on mechanics and those on gas theory) as his "almost morbid ambition".[157]

(b) Boltzmann believed that experimental physicists were at a disadvantage when compared to theoretical physicists because the latter could publish books rooted in their lectures and thereby present theories, quoting Jungnickel and McCormmach,

> …from the perspective of their preferred methods. Boltzmann's published lectures on theoretical physics—covering his favorite parts of it, Maxwell's electromagnetic theory, gas theory, and analytical mechanics—were not syntheses of authoritative writings in the field *but his version of theoretical physics*.[158]

(c) There's evidence that Boltzmann believed that the atomic theory was going to fall out of favor and become completely abandoned. One of his reasons for publishing the *Lectures on Gas Theory* was to produce a historical deposit of the best statement of an atomistic physics of thermodynamics and statistical mechanics (as they pertained to the physics of gases) that he could muster so that when atomic theory was (in his words) "again revived, not too much will have to be rediscovered."[159]

Points (a)-(c) clearly justify a high view of the *Lectures on Gas Theory* understood as the best avenue to Boltzmann's mature thought on statistical mechanics. Interestingly, Boltzmann cites (Boltzmann, On the Relation between the Second Law and Probability Calculus 1877) only once in either of its two volumes. It seemed to have been a theme—not only in Boltzmann's own corpus but also in the work of his contemporaries—that the H-theorem and mechanical approach take precedence.[160] Consider:

(d) Outside of the *Lectures on Gas Theory* and after 1877, there are only five papers/works in which Boltzmann uses the probability calculus, and among these five, only one of them applies the probability calculus to a real-world physical scenario. Among the remaining four papers, two are really just correspondence, and the last two are summaries of his earlier 1877 work.[161]

(e) Boltzmann's combinatorial work was almost entirely ignored by his contemporaries. The standard discussion of the work of both Maxwell and Boltzmann at the end of the 19th century was Rev. Henry William Watson's (1827-1903) *A Treatise on the Kinetic Theory of Gases*.[162] That work never once cites Boltzmann's 1877 paper in which he presents the two combinatorial arguments. Burbury's *A Treatise on the Kinetic Theory of Gases* or (S. Burbury, Treatise 1899) does not discuss Boltzmann's combinatorial approach. Bryan mentions it in a footnote in his contribution to the *Nature* debates. And the principal concern of (Ehrenfest and Ehrenfest 1990) (once an encyclopedia article on statistical mechanics published near the beginning of the 20th century) was the status of Boltzmann's H-theorem.[163]

---

157 As quoted and cited by (Jungnickel and McCormmach 1986, 188).
158 ibid., 189. Except for the word 'his', the emphasis is mine.
159 As quoted and translated by (Jungnickel and McCormmach 1986, 189). See (Boltzmann 1899/1919).
160 See also (M. J. Klein, Mechanical 1973, 73).
161 These points are made by (Kuhn 1978, 70).
162 See (H. Watson 1893). This is the second edition of the work. The first edition was published in 1876.

(f) That Boltzmann's contemporaries understood him to prefer the mechanical approach to thermodynamics and statistical mechanics can be seen in the synopsis of one of his famous students, *viz*., Paul Ehrenfest (1880-1933). He wrote,

> Mechanical representations, were the material from which Boltzmann preferred to fashion his creations…He obviously derived intense aesthetic pleasure from letting his imagination play over a confusion of interrelated motions, forces and reactions until the point was reached where they could actually be grasped. This can be recognized at many points in his lectures on mechanics, on the theory of gases, and especially on electromagnetism. In lectures and seminars Boltzmann was never satisfied with just a purely schematic or analytical characterization of a mechanical model. Its structure and its motion were always pursued to the last detail.[164]

As I've already suggested, Boltzmann published numerous replies as part of a mid-1890s discussion of his work in the journal *Nature*. Discussants included George Hartley Bryan (1864-1928), Burbury, Culverwell, Joseph Larmor (1857-1942), and Watson.[165] That debate took place just after the publication of a new proof[166] of the H-theorem in (H. W. Watson 1893, 33-49). Although no one questioned the correctness of Watson's suitably amended (by Culverwell) proof,[167] many objections and searching questions were raised about how Boltzmann used the H-theorem in his theorizing about irreversible thermodynamic processes and mechanics. In the face of those objections and questions, Boltzmann *never* once abandons the theorem (even though he could have easily reverted to his 1877 combinatorial and probabilistic approach in which the H-theorem played no essential role).[168]

## 7 The Reversibility Paradox Answered

In **sections 1** and **4.1**, I showed that Clausius, Maxwell, and Boltzmann thought of collisions as instances of causation that drive entropic increase (*i.e*., collisions are that which produces the transition from non-equilibrium states to equilibrium states). This fact underwrites the sense in which their way of explaining the second law was mechanical. With respect to Boltzmann and the H-theorem, ensuring minus-H increase requires special types of collisions. Not just any will do. Only collisions with a unique type of built in asymmetry get the job done. I turn now to exploring the full nature of that asymmetry. My exploration will reveal another way in which statistical considerations enter the mechanical explanation of the second law. It will also reveal the solution to the reversibility paradox.

---

163 (M. J. Klein, Paul Ehrenfest 1970, 122).

164 (Ehrenfest, Scientific Papers 1959, 135) as translated by (Klein, Mechanical 1973, 72). *Cf.* (Klein 1974, 166).

165 The series of arguments and replies were published after the August 1894 meeting of the British Association for the Advancement of Science at the University of Oxford. Boltzmann referred to this meeting as "the unforgettable meeting of the British Association at Oxford" (Boltzmann 1964, 22). For many of the details on the discussion I lean, not only on my own readings, but also on those in (Brown, Myrvold and Uffink 2009); (Brush, Vol. 2 1976, 616-625); (Brush 1999); (C. Cercignani 1998, 120-133); (Darrigol 2018, 366-382); (Dias 1994); (M. J. Klein, Ehrenfest 1970, 110-112).

166 The proof had a flaw which Culverwell corrected (Darrigol 2018, 368).

167 Watson thought the proof was purely mechanical.

168 I should add that in (Boltzmann, Certain Questions 1895, 414), Boltzmann does cite his 1877 combinatorial arguments so as to back the claim that he had already argued that the second law of thermodynamics is a statistical law.

### 7.1 The Hypothesis of Molecular Chaos

When in 1895, Boltzmann said that the H-theorem only guarantees that it is highly likely that both (i) appropriate non-equilibrium gas systems increase in entropy over time and (ii) that suitable equilibrium gas systems stay in equilibrium, he said this in reply to the reversibility paradox as articulated, not by Loschmidt but by Culverwell. I rejected Boltzmann's response in **sect. 6.1.1**, because it makes both false and nonsensical claims. I showed, in the same section, that his remarks do not reflect the refined and mature views he communicated elsewhere. Am I preparing the way for a non-statistical statement of the second law? No. As in both the work of Maxwell and Boltzmann, my theory will admit probabilistic considerations in at least two places. First, (again) the Maxwell distribution is itself a statistical principle. Second, it was realized during the *Nature* debates in the mid-1890s that an important assumption—which I will call the hypothesis of molecular chaos (**HMC**)—about the nature of collisions was required in order for the H-theorem to be applicable to real-world systems.[169] With this virtually everyone (whether mathematicians, historians of physics, philosophers of physics, or physicists themselves) agrees. Disagreement arises over the precise form of the assumption.[170] I maintain that the assumption is directly related to my explanation of how and in what way some systems avoid H-theorem application (*q.v.*, **sect. 4.2**). Systems that have very special initial conditions are not guaranteed to be the kind to which the H-theorem is applicable.[171] All positions and velocities consistent with the conservation laws must be allowed early on. One way to *help* ensure that the system does not begin in some special state is to suppose (and Bryan (1895, 29) made this explicit), that the molecular constituents of the system are statistically independent in that their motions are not correlated temporally prior to that which produces entropic increase (*i.e.*, collisions). That is to say, **HMC** states that the pre-collision velocities of two colliding molecules in a gas system of the right kind are statistically independent, and that the post-collision velocities of those same molecules become correlated both after and *because of* the collision. This one-sided or asymmetric molecular chaos propagates for positive times in the sense that collisions that drive minus-H increase retain this correlation-creating ability throughout the system's evolution toward equilibrium. When I say that the velocities after collisions are correlated, I shall at least mean that in order to retrieve the probability of the post-(binary) collision trajectory of one of the molecules

---

[169] See (Bryan 1895, 29); (S. Burbury 1894, 78).

[170] I do not believe the necessary assumption is Burbury's (Condition A) or the Ehrenfests' (1990) Stoßzahlansatz. As I will soon reveal, ***it will not ultimately matter which characterization you choose, for all believe the necessary assumption about the nature of the involved collisions is asymmetric and all believe the assumption is not part of the laws of Hamiltonian mechanics***.

[171] As Villani put it,

> "…for *most* initial configurations, the evolution of the density under the microscopic dynamics is well approximated by the solution to the Boltzmann equation. Of course, this does not rule out the existence of 'unlikely' initial configurations for which the solution of the Boltzmann equation is a very bad approximation of the empirical measure." (Villani 2002, 98)

***It is this idea that Boltzmann's combinatorial arguments are meant to illustrate***.

in the collision, one should conditionalize on the post-(binary) collision trajectory of the other molecule, *inter alia* and *vice versa*.

Bryan was not the first to notice the **HMC** in Boltzmann's project. Something close to it was recognized by Lorentz in his 1886 correspondence with Boltzmann about time-reversal invariance and the derivation of the Maxwell-Boltzmann distribution for polyatomic gas systems. He stated,

> We may assume that in a natural gas the particles have completely *irregular positions and phases*, or at least that there is no definite relation between the positions that the particles have before time *dt* [in which a collision of a given kind occurs] and the number of collisions which they will experience [during this time]. In contrast, it is clear that the positions and phases of the particles are not completely irregular with respect to the past collisions, because they result precisely from the latter collisions. Now, if we revert all velocities as you wish to do, we get a state in which the positions and phases are prepared for the forthcoming collisions and therefore complete irregularity no longer holds.[172]

Here Lorentz articulates the idea that before collisions during *dt*, gas particles are statistically independent and therefore "irregular" with respect to their positions and phases. He likewise affirms that collisions cause those positions and phases to become in some sense regular.

Through some persuasive efforts, Boltzmann came to accept (at least for a substantial period of time) the **HMC**. He wrote, "[w]e shall therefore, [he concludes in 1896] now explicitly make the assumption that the" pre-collision motions are "molecularly disordered and" remain "so throughout all future time."[173] Elsewhere Boltzmann criticizes Gustav Kirchhoff's (1824-1887) derivation of the Maxwell distribution in (Kirchhoff 1894). The basis of his critical review is that Kirchhoff has not properly assumed the **HMC**. Boltzmann turns out to be wrong about this, but the fact that he uses the **HMC** as a criterion for determining the threshold of a good derivation of the Maxwell distribution suggests a high view of the **HMC**.[174]

The more technically inclined reader will desire a formal presentation of the **HMC** in the language of mathematics.[175] I will not provide one because there isn't one. Brilliant mathematicians have given this issue much thought and have concluded that "the physical derivation of the Boltzmann equation is based on the *propagation of one-sided chaos*, but no one knows how this property should be expressed mathematically…"[176] Call this the (**No Mathematics Problem** (**NMP**)). This may strike one as a troubling situation. But matters are worse. The second law is commonly used as part of a solution to the problem of the arrow of time which asks: Why do we perceptually behold irreversible processes when the laws of mechanics governing the micro-constituents of the systems in those processes are time-reversal invariant? If one's answer in any way relies upon the H-theorem, then one's answer will invite yet another problem of asymmetry: Why do the binary collisions that produce minus-H increase produce correlations only *after* those collisions obtain? As Brown et. al. stated, "… there is no reason given as to why the…[**HMC**]…holds for pre-collision velocities rather than post-collision ones."[177] Call this the (**Chaos Asymmetry Problem** (**CAP**)).

---

[172] As quoted and annotated by (Darrigol 2018, 323). Some maintain that Burbury was the first to point out the **HMC** assumption, but this is incorrect. In fact, Burbury required that there be a persisting external perturbation that ensures that systems evolve in a manner consistent with the **HMC** (Burbury 1895, 105). No one accepted Burbury's particular way of couching the **HMC**. Bryan's citation of Burbury in his (1895) work is probably just an attempt to document that the recognition of a related assumption in Boltzmann's work appears in the work of Burbury. Strictly speaking, Bryan's diagnosis of the precise content of the assumption was different from the content of Burbury's diagnosis.

[173] (Boltzmann 1964, 42). See also ibid., 58-59; (Boltzmann, Maxwell's Distribution Again 1895); (C. Cercignani, Boltzmann 1998, 259); and (Kuhn 1978, 64). Something like the idea expressed here may even be in the work of Clausius (Ehrenfest and Ehrenfest 1990, 5).

[174] For more on the Boltzmann-Kirchhoff debate, see (Darrigol 2018, 320-321; 360-361).

[175] Sometimes the assumption is said to be equivalent to the claim that the distribution function satisfies: $f^{(2)}(\mathbf{v}_1, \mathbf{v}_2) = f(\mathbf{v}_1)f(\mathbf{v}_2)$, where $f^{(2)}$ is the distribution function for a pair of molecules. Here the idea is that the probability of seeing a pair of molecules with velocities $\mathbf{v}_1$ and $\mathbf{v}_2$ (around $d^3\mathbf{v}_1$ and $d^3\mathbf{v}_2$ respectively) is equal to the product of finding a molecule with $\mathbf{v}_1$ around $d^3\mathbf{v}_1$, and a molecule with $\mathbf{v}_2$ around $d^3\mathbf{v}_2$ (Callender 2011, 85); *cf.* the remarks at (Uffink 2007, 1036) on the BBGKY approach. Villani (2002, 99) argues that this is not an adequate characterization and that it actually needs to be generalized sufficiently to get the right result. Villani does not know how to do this and worries that it can't be done. I agree with Villani and rest on his authority.

### 7.2 The Solution at Long Last

My proposed resolution of the reversibility paradox will also serve as the solution to the **NMP**, and the **CAP**. The first step of the solution is to understand the **HMC** as an interpretive postulate about the nature of that which drives minus-H increase, *viz*., collisions. That the collisions are responsible for entropic increase in Boltzmann's H-theorem-laden kinetic theory is acknowledged by virtually all scholars.[178] The standard story in kinetic theory is that collisions between molecules in non-equilibrium closed systems drive those systems into equilibrium, and that "equilibrium" writes Thomas Kuhn, "is, by definition, the state in which the distribution is unaffected by collisions."[179] But now we must ask, if collisions causally produce minus-H increase, then why do collisions among molecules of gas systems in equilibrium fail to increase minus-H even further? Of course, once the system reaches equilibrium it is characterized by the Maxwell distribution, in which case, the functional H vanishes thereby reaching its minimum value (the H-theorem used to be called the minimum theorem). It's just a mathematical fact that H cannot decrease, and that minus-H cannot increase. But mathematical facts can have metaphysical explanations. That is to say, there exists a reason why once H vanishes, entropy fails to increase, and that reason consists in the fact of energy dissipation. Recall Stephen G. Brush's point (quoted previously) that "the H-theorem is a microscopic version of the general principle of [the] *dissipation of energy* proposed by Kelvin in 1852, and reformulated by Clausius in 1865...'"[180]

---

[176] (Villani 2002, 99). In his well-regarded book, Herbert Spohn remarked, "…the decrease of [the] H-function is linked to instants of molecular chaos. *These properties remain a guess*." (Spohn 1991, 76) emphasis mine.

[177] (Brown, Myrvold, and Uffink 2009, 181). See the same point in (Price 1996, 40).

[178] See also (Brown, Myrvold and Uffink 2009, 175); (Brush, Vol. 2 1976, 443-444, "the later Maxwell-Boltzmann developments [are] based on consideration of molecular collisions" 619); (Brush 1999, 25-26 "Boltzmann proved that…collisions always push f(x,v,t) toward the equilibrium Maxwell distribution" ibid. and see ibid., 22 on the idea in Maxwell); (Callender 2011, 89 reporting in n. 3 that Jos Uffink agrees); (Darrigol 2018, 321-323 on the idea in Lorentz's thought); the idea is clearly in related work by Kirchhoff, for which see (ibid., 361); (Jungnickel and McCormmach 1986, 64, with remarks about Boltzmann); (M. J. Klein, Ehrenfest 1970, 100, attributing the view to Boltzmann, see also p. 102); (Maxwell 1867, 62, 64); (Segrè 1984, 279); (Sklar 1993, 32).

[179] (Kuhn 1978, 62).

[180] (Brush vol.1 1976, 80). It should not surprise us then to see in Boltzmann's interpretation of the second law as explained by the H-theorem, remnants of Thomson's (and Clausius's) idea of energy dissipation. Those remnants show up in Maxwell's own interpretation (which influenced Boltzmann's work) of the second law as well. Although energy dissipation in Maxwell's thought possessed a certain anthropocentric element. See (Smith 2003, 303-304 and n. 41); (Smith 1998, 240-241; 247-252); (Smith and Wise 1989, 623). For Maxwell's actual work, see *SPM2*, 646.

The way energy has transformed and dissipated—remember entropy tracks this energy transformation—has left the system in equilibrium no longer allowing it to further transform.[181] The capability of the system to perform work becomes attenuated. Contemporary thermodynamicists such as Sanford Klein and Gregory Nellis interpret the second law "as a system for assigning quality to energy." They continued,

> Although energy is conserved, the quality of energy is always reduced during energy transformation processes. Lower quality energy is less useful to us in the sense that its capability for doing work has been diminished. The quality of energy is continuously degraded by all real processes; this observation can be expressed in lay terms as 'running out of energy'.[182]

The energy transformative process is a causal one. That interpretation is plausible for at least two reasons. First, at the heart of the process in thermodynamic or statistical mechanical evolutions are causally efficacious collisions producing entropic increase. Second, kinetic energy is a causal quantity. Rankine said that "actual", or what, in 1862, Thomson would identify as kinetic energy "is a measurable, transferable, and transformable affection of a substance, the presence of which causes the substance to tend to change its state in one or more respects…".[183] Modern statements do not differ, as contemporary classical (non-relativistic) physics universally characterizes kinetic energy in terms of work. Changes in kinetic energy $dT$ (where $T$ is not temperature but kinetic energy) are also specified by appeal to work done by net force, or $\mathbf{F} \cdot d\mathbf{r}$.[184] But forces in classical mechanics are causal.[185]

---

[181] There are worries about Poincaré recurrence and fluctuations looming. I have answers for those worries too. My explication of them must be left for another project.

[182] (Klein and Nellis 2012, 2 *cf.*, 350). We do have to be careful not to mix up or confuse energy and exergy. Exergy is also a useful quantity in thermodynamics. It is defined as "the capability to do useful work" (ibid., 351).

[183] (Rankine 1853, 106).

[184] In the case of disagreeing angles one affirms: $W = \mathbf{F} \cdot s\ \cos\theta$.

The rate of changes of kinetic energy are what's important, for $T = \frac{1}{2}mv^2$ (for the single classical point mass) never has an absolute value because the point mass's velocity or speed will be relative to a reference frame.

[185] This was the opinion of Newton, Leibniz, Huygens, Lagrange, Hamilton, Laplace, Maxwell, Boltzmann, Helmholtz, Gibbs and a great many others. I'll *very* briefly focus on Newton and Hamilton because they are the most relevant in this context.

Newton: Newton said that "forces…are the causes and effects of true motions." (Newton 1999, 414). The entire purpose of the *Principia* is given in this statement at the end of the *Scholium*:

> "But in what follows, a fuller explanation will be given of how to determine true motions from their causes, effects, and apparent differences, and, conversely, of how to determine from motions, whether true or apparent, their causes and effects. For this was the purpose for which I composed the following treatise." (Newton 1999, 415)

Hamilton: Sir William Rowan Hamilton's (1805-1865) causal mechanics was indebted to Immanuel Kant's (1724-1804) "Second Analogy of Experience" in the first *Critique* (Kant 1998, 304-316). Like Kant, Hamilton believed that every dynamical evolution had to involve some causality (Hankins 1980, 179). In the first of Hamilton's two most famous papers on dynamics, "On a General Method in Dynamics" (Hamilton 1834)/(Hamilton 1940, 103-161), Hamilton reasons to what he calls the **law of varying action** (**LVA**):

(Eq. 1 n. 185):

$$\delta V = \sum m\,(\dot{x}\delta x + \dot{y}\delta y + \dot{z}\delta z) - \sum m\,(\dot{a}\delta a + \dot{b}\delta b + \dot{c}\delta c) + t\delta H$$

Some think we can forsake forces in a conceptually sophisticated enough classical (non-relativistic) mechanics if we appropriate Hamiltonian mechanics, an energy-based theory. Hamiltonian mechanics is an energy-based approach to the dynamics of classical (non-relativistic) systems because the laws of motion in Hamiltonian mechanics use the Hamiltonian or energy function (and here, I follow Taylor 2005, 528-531):

(15):

$$\mathcal{H} = \sum_{i=1}^{n} (p_i \dot{q}_i) - \mathcal{L}$$

given that $[i = 1, \dots, n]$, and that the system is described by generalized momenta:

(16):

$$p_i = \frac{\partial \mathcal{L}}{\partial \dot{q}_i}$$

and specified by generalized coordinates:

(17):

$$\mathbf{q} = (q_1, \dots, q_n)$$

Here the Lagrangian $\mathcal{L}$ is a function of $\mathbf{q}, \dot{\mathbf{q}}$ (specified below), and time. If the system is isolated, the generalized coordinates stand in a time-independent relationship to the Cartesian or rectangular coordinates tracking the system, and the potential energy of the system is velocity independent,

(18):

$$\mathcal{H} = T + U$$

and generalized momenta as well as generalized velocity can be written (respectively) as:

(19), (20):

$$\mathbf{p} = (p_1, \dots, p_n), \quad \dot{\mathbf{q}} = (\dot{q}_1, \dots, \dot{q}_n)$$

---

also calling it the "equation of the characteristic function" $V$ (Hamilton 1834, 252). $V$ "completely determines the mechanical system and gives us its state at any future time once the initial conditions are specified" (Hankins 1980, 186). At the time, the function $V$ was sometimes called the action of the system, hence "law of varying **action**".

The above statement of the **LVA** entails that $V$ is a function of the $3n$-coordinates for whatever point masses are in the system, and the Hamiltonian $H$. As I point out in the main text above, for conservative systems:

(Eq. 2 n. 185):

$$H = T + U$$

Kinetic and potential energy enter the **LVA** through $H$. Importantly, Hamilton calls $U$ the force-function because it is always associated with a corresponding force (Hamilton 1834, 249). In addition, Hamilton explicitly connects variations of $U$ to work done by subsystems (Hankins 1980, 184), and also defines $U$ in terms of a force law (Hamilton 1834, 249). For Hamilton, this is how dynamics is causation-laden.

Both potential and kinetic energy are analyzed (at least in part) in terms of work (force times displacement). When the Hamiltonian equals kinetic and potential energy, force thereby enters Hamiltonian mechanics. When it is appropriate to specify the Hamiltonian in terms of the Lagrangian $\mathcal{L}$, force enters more indirectly.[186] The Lagrangian $\mathcal{L}$ is (in many appropriate circumstances) equal to $T - U$ (where $U$ is here potential energy and not internal energy). So, the Lagrangian is (in appropriate circumstances) at least in part specified by appeal to the kinetic and potential energy of the system. But again, kinetic and potential energy, even in Hamiltonian mechanics, is, in part, standardly interpreted and analyzed in terms of work. But work is, in part, specified in terms of net force. Thus, forces are indispensable to any plausible interpretation of Hamiltonian mechanics, and therefore causation is as well since forces are causes (*Hamilton would agree*! *Q.v.*, n. 185).

If our interpretation of classical mechanics is causal, then it admits an asymmetry. Causation is formally asymmetric. How then do I meet the famous reversibility objection in the work of Thomson, Loschmidt, and Culverwell? Recall the gist of that objection. All minus-H increasing evolutions imply the possibility of minus-H decreasing reversed evolutions of an appropriate isolated gas system. This, thought Thomson, Loschmidt and Culverwell, is a consequence of the reversibility of the microdynamics (Darrigol and Renn 2013, 774). The reversibility of the microdynamics and therefore also the reversibility of the supervening macroscopic evolutions was thought to be a consequence of the reversal of the involved velocities. As Loschmidt wrote, "the entire course of events will be retraced if, at some instant, the velocities of all its parts are reversed."[187] Or as Thomson put it, "[i]f, then, the motion of every particle of matter in the universe were precisely reversed at any instant, the course of nature would be simply reversed for ever after."[188] Reversing velocities was deemed naturally possible because the underlying microdynamical equations of motion in Hamiltonian mechanics were correctly thought to be time-reversal invariant.

The idea that you get so much from simple *velocity* reversal of microconstituents of *real-world* classical systems is mistaken. Entropic increase as envisioned by the H-theorem-laden kinetic theory is not driven by an underlying microdynamical evolutionary *process* that is reversible. Am I denying that Hamilton's equations of motion are time-reversal invariant? ***No***. Recall that those equations are time-reversal invariant only if replacing *t* with minus-*t* (being careful to also flip the sign of all odd forms of *t* such as velocity) allows solutions to be taken to solutions (or as Thomson said, "any solution remains a solution" (1874, 441)). I am certainly not denying that. It's a mathematical fact. However, the equations of motion, *once fully interpreted and thereby rendered applicable to real-world classical systems*, inform us about unfolding causal processes that possess an asymmetry even in the micro-processes. That asymmetry stems from the causation in collisions, the very engine of entropic increase and so also the source of the asymmetry in the **HMC**. This source is not *directly* represented by the mathematics expressing the microdynamical (Hamiltonian) laws and so it is no surprise that the **HMC** is not directly represented in that mathematics either. The **HMC** is not part of the formulation of mechanical principles that govern collisions (compare Uffink's remarks in 2007, 969). Rather, it is an understanding of how that formalism fits the real world (*i.e.*, it is part of an interpretation of the mechanics).[189] But one might counter: The **HMC** is about collisions, and we have a classical mathematical collision theory.

[186] And I do have in mind the Lagrangian and *not* the Lagrangian density.
[187] (Loschmidt 1876, 139) as quoted and translated by (C. Cercignani 1998, 98).
[188] (Thomson 1874, 442).

### 7.2.1 Solving the Chaos Asymmetry and No Mathematics Problems

Go back to Maxwell's "On the Dynamical Theory of Gases" (1867). There, Maxwell assumes that all collisions are elastic (total kinetic energy and momentum are conserved through collisions). This assumption is false for polyatomic molecules, and false for atomic collisions. The latter conjunct holds because in collisions between atoms some kinetic energy is converted into other forms of energy. But set these points aside. As in some modern accounts of classical collision theory, Maxwell accounted for constituent collisions between two arbitrary gas molecules by giving attention to their pre-collision velocities, their post-collision velocities, and those collision "parameters that are necessary to determine the final velocities of the molecules."[190] As already revealed in preceding discussion (*q.v.*, n. 39), the collision parameters are usually the azimuthal angle $\phi$, and the impact parameter $b$. With respect to *binary collisions*, the latter is nothing more than two modal entities, *viz*., the paths the two molecules would travel were they to fail to *interact* with one another (in the center-of-mass frame). The former is just an angle, *viz*., the angle that fixes the plane upon which sits the post-collision trajectories of both molecules. The type of interaction involved need not be restricted to a physical contact in a real-world collision because the types of entities interacting are not restricted to or always best approximated by point masses. The interaction may be complex involving various force-types. For example, it may include the exertion of non-contact forces made manifest in attractions or repulsions alongside or with contact forces. But even if the involved force impressions were purely contact forces, the impact parameter would not provide that which is sufficient for fully determining (in the sense of producing) the post-collision velocities. For elastic collisions of molecules of gases (not unlike ideal gases) with the same masses, one can through straightforward mathematical reasoning, determine (in the (epistemic) sense that you can infer or *derive*) the post-collision velocities of the two colliding molecules from knowledge of the laws of conservation, the impact parameter, the azimuthal angle, and the pre-collision velocities. The sense of determination here is epistemic because it would be obviously shortsighted to judge that because a mathematical fact about the post-collision velocities follows from mathematical facts about conservation, the pre-collision velocities, the azimuthal angle, and the impact parameter, that therefore nothing more in the world *metaphysically* determines the post-collision velocities when the phenomenon under study is a *collision phenomenon* involving impact force impression. There was a collision! There was an interaction between the two molecules! What has happened in Maxwell's treatment (and as we shall see in Boltzmann's treatment too) is that Maxwell has chosen to model around the interactions, or the intimate details of the impact-laden collisions.[191] What has happened is that Maxwell has utilized

---

[189] On this distinction, see (Weaver 2019, 52-71) and the literature cited therein.

[190] (Darrigol 2018, 82). A modern account resembling Maxwell's can be found in (Taylor 2005, 557-593).

[191] You see this in the way he characterizes collisions. Writing to Stokes in 1859, he said,

> "I saw in the Philosophical Magazine…a paper by Clausius on the 'mean length of path of a particle of air or gas…'…on the hypothesis of the elasticity of gas being due to the velocity of its particles and of their paths being rectilinear except when they come into close proximity to each other, *which event may be called a collision*." (Maxwell vol.1 1990, 606 emphasis mine)

In his 1867 paper he writes,

a conceptual strategy that Mark Wilson calls *physics avoidance* (Wilson 2017). Maxwell sought to model the world using quantitative walk-arounds that enabled his models to escape severe mathematical difficulties.[192]

---

> "In the present paper I propose to consider the molecules of a gas, not as elastic spheres of definite radius, but as small bodies or groups of smaller molecules repelling one another with a force whose direction always passes *very nearly through the centres of gravity of the molecules*." (Maxwell 1867, emphasis mine)

Quite clearly Maxwell had in mind molecules that interact in other ways besides elastic collisions involving impacts. But as I stated in the main text, real world molecules *and* particles interact by means of repulsions or attractions *plus* impacts. For example, there are electron-electron collisions or scatterings, especially at high energy levels, despite coulombic repulsion (Lee et. al. 2020). In dense plasma recombination phenomena, electron-electron collisions occur. However, these recombinations are not similar to ionic three-body recombination phenomena precisely because of operating Coulombic forces in the former recombination cases (Bates and Kingston 1961). My reader will retort that the molecular or particulate world is a quantum world. Sure. But in the phenomenon of ionization as causally produced by a free electron, the free electron comes in and strikes, thereby impacting, an electron bound to an atom. The energy transferred to the bound electron is greater than the binding energy of the bound electron. Thus, the impact and resulting energy transfer frees the bound electron from the atom. It is true that such a case is captured or explained by quantum physics, however, the scattering involved is elastic and the cross-sections of each electron are the same in both the quantum and classical domains (Kosarim et. al. 2005, 215. The electron-electron interactions they discuss are cases involving real impact. See the very title of their paper.); *cf*. (McCarthy and Weigold 1990). You can therefore "use classical methods for [the] evaluation of the ionization cross-sections of an atomic particle *by electron impact*" (Kosarim et. al. 2005, 215 emphasis mine). I can ensure the relevance and accuracy of classical physics for this phenomenon by restricting my discussion to slower electron velocities and non-highly excited atoms. I do this because Hans Bethe (1906-2005) (this point is made by ibid.) showed that with respect to large electron velocities, an additional (beyond the classical) logarithmic factor exists in the cross-section of ionization (Bethe 1930). The classical method used by Kosarim et. al. adequately accounts for the experimental data.

It is sometimes said that the molecules of ideal gases do not interact at all (Frigg 2008, 119). That is not true. The equation of state for ideal gases (*i.e*., the ideal gas law) includes the quantity that is pressure. Pressure is force over unit area. If the ideal gas were confined to a container, the molecules would causally produce pressure by interacting with or impacting the boundaries, themselves atomically constituted, of that container. There would fail to exist pressure in the system if there were no such interactions. This is why modern work in thermodynamics assumes that ideal gas molecules do in fact undergo interactions with perfectly elastic and adiabatic boundaries. In fact, ideal particles or molecules can bring about "irreversible work contributions" through transferring momentum with a moving piston by interacting with that piston (Hanel and Jizba 2020, 2, 13). The types of interactions that are precluded in the ideal gas case are interactions via repulsions and attractions. How else could an ideal gas reach thermal equilibrium if its velocities never changed as a result of accelerations wrought by impressed (at least impact) forces? Modern theorists are careful to note that "[f]or an ideal gas interactions between all molecules are supposed negligible, *other than for establishing thermal equilibrium*" (Bowler 2017, 3 emphasis mine). That ideal gas constituents collide *with each other* thereby impressing impact forces upon each other, is the standard view (Argo 1981, 25; Keeton 2014, 244; Wilson 1994, 351, citations could be multiplied).

[192] I should add that modeling from a distance is also important to Maxwell because when many molecules collide matters become intractable. This is not because we lack the ingenuity to solve the equations appropriately, *it is because we do not have the right equations*! He wrote, "[w]hen we come to deal with collisions among bodies of unknown number, size, and shape, *we can no longer trace the mathematical laws of their motion* with *any* distinctness". (Maxwell 1859, 53) emphasis mine; (*SPM1*, 354). Garber adds,

> "He [Maxwell] concluded by noting *the inability of dynamics* to address this last problem…Mechanics cannot deal with collisions among many bodies flying around…." (Garber 2008, 1701) emphasis mine

Boltzmann's proof of the H-theorem treats collisions much the same way Maxwell modeled them, *i.e.*, by using the conservation laws, plus the "initial value of the kinetic energies $k_1$ and $k_2$ of" two colliding molecules, "and by the value $k'_1$ of the kinetic energy of the first molecule after the collision."[193] Boltzmann's combinatorial argument of 1877 practices a similar type of physics avoidance, except in that context it completely "neglects the contribution to the energy of the system that stems from interactions between the particles."[194] Followers of Boltzmann who prefer the combinatorial method do not resist the neglect. In fact, the same Boltzmannians provide a means whereby we can empirically distinguish H-theorem-laden statistical mechanics from the more popular modern Boltzmannian approach found in places like (Albert 2015), (Goldstein and Lebowitz 2004), (Goldstein, Tumulka and Zanghì 2016), (Lebowitz and Maes 2003), and (Loewer 2008). For example, Goldstein et. al. call the entropy discussed in Boltzmann's combinatorial approach, Boltzmann entropy ($S_B$). They provide sound justification for separating $S_B$ from the entropy that is minus-H, stating that

> *[i]f interaction cannot be ignored*, then the *H* functional does not correspond to the Boltzmann entropy…*[w]hen interaction can be ignored* there is only kinetic energy, so the Boltzmann macro states based on the empirical distribution alone determine the energy and hence the H functional corresponds to the Boltzmann entropy.[195]

In modern classical mechanical approaches to Boltzmannian statistical mechanics that use an H-theorem and a Boltzmann collision operator $Q$, impact interactions are avoided or modeled around (as is implied in (Villani 2002, 79)). In that context too, collisions are all assumed to be binary, and the involved particles don't really contact one another, for in that literature binary collisions are processes "in which two particles happen to come very close to each other, so that their respective trajectories are strongly deviated in a very short time."[196] There is physics avoidance afoot here because modern theoreticians are engaging in modeling walk-arounds.

Ignoring interactive contact collisions between point-like objects is as old as Newtonian mechanics. Newton's second law says that "[a] change in motion [not a rate] is proportional to the motive force impressed [where the proportionality constant is inertial mass] and takes place along the straight line in which that force is impressed" (Newton 1999, 416).[197] The masses of the objects to which the second law was intended to apply never equal zero. Even point masses have mass.

---

For Clausius, collisions and even "impacts" resulting in rebound effects are not instances in which centers of gravity or gas constituents literally come into contact with one another. It was enough for Clausius that the centers enter one another's spheres of action (*q.v.*, my discussion of Clausius in **sect. 1** above).

[193] (Darrigol 2018, 139). Boltzmann wrote,

> "Das Produkt dieser drei Größen muß noch multipliziert werden mit einem gewissen Proportionalitätsfaktor, von dem man leicht einsieht, daß er unendlich klein, wie $d\xi$ sein muß. Derselbe wird im Allgemeinen von der Natur des Zusammenstoßes, also von den, *den Zusammenstoß bestimmenden Größen x, $x'$ und $\xi$ abhängen*." *BWA*1, 324 emphasis mine

Here Boltzmann clearly states that the nature of the binary collisions is determined by pre-collision kinetic energies and the one post-collision kinetic energy.

[194] (Frigg and Werndl, Boltzmann Equation forthcoming, 6).

[195] (S. Goldstein, et al. 2019, 28).

[196] (Villani 2002, 79).

[197] The best discussion of how Newton understood his second law of motion can be found in (Pourciau 2006), although I would add and emphasize a causal force ontology in Newton's thought.

However, when two point-like objects or point masses collide, their accelerations are obliterated, and as a result the second law fails as there is a force impressed but no resulting acceleration (and not because of a balance rendering the force vector equal to the zero vector). Newton was aware of this problem and saw no application of the second law of motion to contact interactions. Here is Wilson on Newton's approach to the problem. Note the similarities to the methods of Maxwell and Boltzmann,

> Whenever these radii contact one another (we shall only worry about the head on collision case), Newton abandons the requirement that the '**a**' in '**F** = m**a**' must make sense and shifts his focus to the two balls' incoming stores of linear momentum and kinetic energy (as we now dub them), together with a purely empirical factor called a *coefficient of restitution* (it governs how much the total kinetic energy budget will diminish post-collision). In effect, this treatment blocks out the crucial interval of time $\Delta t$ where '**F** = m**a**' fails to make sense and glues together the incoming and outgoing events exterior to $\Delta t$ through a mixture of conservation principles (conservation of linear momentum) and raw empirics (coefficients of restitution extracted from experiment). Formally, tactics that patch over problematic intervals or regions in this manner are frequently called *matched asymptotics*.[198]

The problem is not unique to Newton's formulation of classical mechanics. It reappears in Hamiltonian mechanics. Mathematician Robert Devaney stated,

> …specific Hamiltonian systems which arise in applications often suffer singularities as well. By a singularity we mean a point where the differential equation itself is undefined. A typical example of a singularity is a collision between two or more of the point masses in the Newtonian *n*-body problem. At collision, the differential equation breaks down: the velocities of the particles involved become undefined. A singularity or collision can create havoc among nearby solution curves. Solutions which pass near a singularity may behave in an erratic or unstable manner, and solutions which start out close to one another can end up far apart after passing by a singularity.[199]

That you should care about more than mere positions and post-collision velocities in such cases, and that you should give attention to the interactions during the relevant $\Delta t$ (sometimes this time interval is referred to with the symbol $\Delta t^*$) was expressed very clearly by Gottfried Wilhelm Leibniz (1646-1716). The fact that you could recover so much while ignoring the details of the

---

[198] (Wilson 2013, 69). Some will object. They will note that if Newton's *Principia* does anything it provides the correct physics of billiard ball interactions and evolutions. This is not the case (Wilson 2006, 567-598). As Wilson has said,

> "What should be properly said is that Newton and his followers practiced an admirable restraint in their descriptive ambitions, by substituting a crude but reliable *walk-around method* for a very difficult moving boundary computation. *Even today, modern models of impact follow a Newtonian pattern whenever they can get away with it…*" (Wilson 2017, 105. n. 13 emphasis mine) evolution during $\Delta t$ was, for Leibniz, "a convenient trick".[200] Leibniz thought that in order to get to the deep joints of nature, you need to pick up on what's transpiring during the relevant $\Delta t$. Like Leibniz, I maintain that what one will find there (*i.e.*, in the relevant $\Delta t$) is efficient causation or causal interaction (Leibniz 1998, 139-142).[201] *That causation* (call it **fundamental causation or causation$_F$**) drives the engine of entropic or minus-H increase. It results in correlations (that's why you can use correlations to find causal interactions), correlations that are one-sided precisely because **causation$_F$** is asymmetric. That is to say, obtaining **causal$_F$** relations in entropy producing collisions explain the **HMC**. The **Chaos Asymmetry Problem** (**CAP**) has been resolved. The propagating one-sided chaos referenced by the **HMC** is one-sided because the velocity correlations are the effects of temporarily prior causes in temporally directed obtaining **causal$_F$** relations. It is no surprise then that the Boltzmann equation breaks T-symmetry. It does this because the collisions it is about involve obtaining **causal$_F$** relations that are temporally asymmetric.

[199] (Devaney 1982, 535).

The introduction of **causation$_F$** into entropy increasing collisions during the relevant $\Delta t$s resolves the **No Mathematics Problem** (**NMP**) as well. There's no mathematical representation of the **HMC** because its source is unrepresented by the formalism and because the correlations **HMC** references are set down during $\Delta t$. Our best modeling of the collision process walks around those times since its chief concern is recovering post-collision velocities. We can nonetheless point to that best modeling as evidence of the existence of **causation$_F$** in the $\Delta t$s because that modeling, while one step removed from the phenomena, nonetheless recognizes that forces and resulting accelerations obtain so as to get the velocity changes. Applying the time-reversal invariance operation will not change the directions of the forces (the causal structure) nor the directions of the resulting accelerations (though the displacement is reversed). Forces and accelerations are even forms of *t*.

### 7.2.2 Solving the Reversibility Paradox

To see the resolution of the reversibility paradox, recognize first that the time-reversal invariance operation in Hamiltonian mechanics is one that is applied to Hamilton's equations of motion (and appropriate deductive consequences thereof). Odd forms of *t* receive sign changes and solutions are still mapped to solutions. Execution of that operation together with the execution of an appropriate time-translation (so as to help us appreciate a temporally reversed evolution) *will not* entail that there exists an evolution that satisfies a temporally reversed **HMC**. This is because the **HMC** is not a part of the formulation of Hamiltonian mechanics, nor is it a deductive consequence of the equations of motion in Hamiltonian mechanics. Again, **HMC** is an interpretive hypothesis.

---

[200] (Wilson 2017, 116). See (Leibniz 1989, 124). Wilson goes on to point out that Leibniz was at the time concerned with a cut-off method employed by Christiaan Huygens (1629-1695). That cut-off procedure resembles the matched asymptotics of both Newton and modern modeling.

[201] As Wilson's summary of Leibniz stated,

> "…it is only by plowing over these $\Delta t^*$ events that we can explain the elastic behavior of our original wooded beam in a purist efficient causation manner that speaks of nothing but the pushing and pulling of contacting particles." (Wilson 2017, 117)

I should add that unlike Leibniz I see no room in the temporal intervals for the final causation that is discussed in the context of detailing the importance of "the mutual interactions of bodies" (Leibniz 1998, 142).

Suppose there's an elastic impact collision C between two molecules (1 and 2) of a monatomic gas system. Molecules 1 & 2 have velocities $\mathbf{v}_1$ and $\mathbf{v}_2$ (respectively) *before* C. *After* C, they take velocities $\mathbf{u}_1$ and $\mathbf{u}_2$ (respectively). $\mathbf{v}_1$ does not equal $\mathbf{u}_1$, and $\mathbf{v}_2$ does not equal $\mathbf{u}_2$, and I will assume that molecule 1's mass is larger than molecule 2's inertial mass, but not significantly larger. Velocities $\mathbf{v}_1$ and $\mathbf{v}_2$ produce C. C produces post-collision velocities $\mathbf{u}_1$ and $\mathbf{u}_2$. The fact that the accelerations and force impressions in C are not reversed under time-reversal suggests that in the ***time-reversed evolution***, the velocity transitions run from $-\mathbf{u}_1$ and $-\mathbf{u}_2$ to $-\mathbf{v}_1$ and $-\mathbf{v}_2$. So, in the reversed evolution, you won't approach the Maxwell distribution precisely because the velocity transitions/changes go in the wrong direction. They go in the wrong direction because of the fundamental causal structure of the evolution. The (one-step-removed) evidence for this resides in the fact that in the reversed evolution, the forces are still pushing in the same directions as the actual world evolution, and the accelerations keep their actual world directions as well. Because the **HMC** is an interpretive postulate, the time-reversal operation alone will not change its one-sidedness either. Whatever is done with the **HMC** in the reversed evolution is done by hand. *The causal structure of the world must be changed to realize reversed evolutions*.[202]

What of the classical possible world *w* at which monatomic gas systems of the right kind evolve in perfect accord with the models of Maxwell (1867) and Boltzmann (1872, 1875)? At *w*, will there fail to be mon*otonic* increase of $-H$, if such gas systems begin their evolutions in low entropy states? At *w*, all binary "collisions" never introduce problems of mathematical singularities because the constituent molecules never meet. My project seeks to causally *interpret* only those collisions quantified over by the **HMC**. My central thesis, **CC** in **sect. 0**, made this clear. What I'm recommending is that we understand the **HMC** as an interpretive postulate about the types of collisions that transpire during the crucial $\Delta ts$. As I've argued, both Maxwell and Boltzmann did not include specific reference to such impact collisions in their mathematical models because (on my interpretation) they were employing the conceptual strategy of physics avoidance (hence the **NMP**). They were able to discover the various distribution laws because the model walk-arounds "do the trick" (as Leibniz would say) of recovering the velocities of gas molecules after the collisions that are walked around.[203] They encounter a reversibility paradox precisely because the surface meaning of their modeling describes systems like those in *w*, systems whose dynamical evolutions are such that their time-reversal yields a *past*-directed evolution. In *w*, $-H$ does not monotonically increase in accordance with the H-theorem. How could it? The evolutions there are completely time-reversable. Nonetheless, there is no violation of the H-theorem there because the **HMC** (a precondition of the theorem) fails to hold at *w*. The collisions the **HMC** quantifies over

[202] If you follow the many philosophers of physics who maintain that the crucial asymmetric assumption of Boltzmann's reasoning is different from the **HMC** as I have stated it, and that it is, instead something like the Stoßzahlansatz as explicated by the Ehrenfests, then you would do well to note that in (Ehrenfest and Ehrenfest 1990, 85. n. 65) a proof-sketch is summarized. The argument shows that the Stoßzahlansatz *cannot* hold in both the real world and reversed evolutions. Compare the similar stronger argumentation in (Burbury 1895, 320 I skip the meat and potatoes and give the thesis and conclusion),

> "I said in my first letter on this subject that the condition A [an asymmetric assumption like the HMC], on which, or its equivalent, the proof is based, could not apply to the reversed motion. As that assertion has been questioned, may I confirm it thus?...Boltzmann's theorem can be applied to both motions only on condition that it has no effect in either."

[203] Why are they able to do the trick? How can time-reversal invariant modeling, modeling which when reversed yields *past*-directed evolutions, recover descriptions of asymmetric *future*-directed evolutions? That is a very interesting question, a question which Leibniz believed suggested teleology. I will not delve into this particular matter.

do not transpire there and so neither does entropic increase of the kind *required by the H-theorem*. But as soon as we shift to the real world, where monatomic gases like helium (**He**), argon (**Ar**), xenon (**Xe**) and others, evolve in ways featuring real world impact collisions avoided by the Maxwell-Boltzmann modeling *but targeted from afar by that modeling nonetheless* (and so my project remains true to the spirit of Boltzmann's work), the **HMC** becomes part of a true and correct (in the appropriate limit) interpretation of Hamiltonian mechanics being made true by the causal structure of the actual world. It is the *contingent* causal way the world is that determines the entropic asymmetry described by the H-theorem. It is a consequence of my framework that the proposed interpretation of Hamiltonian mechanics makes a detectable empirical difference. It is to that empirical difference that I now turn.

Is the **HMC** empirically justified? Yes. It is indirectly justified by all the fruit or empirical success produced by the H-theorem *and* Boltzmann equation in modern kinetic theory. For example, it should be obvious by now that the H-theorem predicts that if a classical monatomic gas system *SYS* satisfies certain conditions, then *SYS* will evolve to thermal equilibrium over a sufficiently long period of time. That is in fact what we observe. More generally, the H-theorem predicts the truth of the second law of thermodynamics for systems that satisfy the antecedent of the theorem. Consequently, in a restricted sense, the theorem "demonstrates the second law of thermodynamics."[204] Nature's obedience to the second law is what we observe. In addition, I have already indicated how the Boltzmann equation is used to great benefit in the study of neutron transport, plasma physics, and the kinetic theory of gases (*q.v.*, **sect. 3**; and see Cercignani 1988; Cohen and Thirring 1973; White et. al. 2009). What is more, the H-theorem and Boltzmann equation bear much fruit in hydrodynamics as well (Succi, Karlin, and Chen 2002). The empirical successes of the Maxwell and Maxwell-Boltzmann distributions discussed in the sources at note 43 are also relevant indirect justifications of the **HMC**. Why believe the above constitutes indirect evidence for the **HMC**? The **HMC** "is a fundamental requirement for the application of the Boltzmann kinetic theory, the Boltzmann transport equation, and the presence of Maxwell-Boltzmann statistics."[205]

There is a recent and more direct justification as well. I call this other justification "more direct" and not "direct" *tout court* because we are not currently able to directly observe the correlated velocities of gas molecules. However, there are a class of granular media that are low-density media which approximate gas systems (they are called "granular gases" in light of this). G.W. Baxter and J.S. Olafsen gave attention to such systems in 2007. They discovered that these low-density granular systems exhibit molecular chaos, but that once the systems become sufficiently dense (*i.e.,* once there are sufficient enough interactions (this is my gloss)), the velocities of the constituents of the relevant systems become correlated. [206]

[204] (Gressman and Strain 2011, 2351).

[205] (Baxter and Olafsen 2007, 1). See also (Huang 1987).

[206] They remarked,

> "The relative lack of velocity correlations in the second layer at low densities is evidence of the presence of molecular chaos in this system. The upper layer continues to demonstrate uncorrelated velocities until the density reaches 80%." (Baxter and Olafsen 2007, 4).

They would add that they are unsure of how it is precisely that the correlations obtain in the system, but it seems clear that the interactions play a key role. Why else would density matter?

## 8 Conclusion

I have shown that the Standard Story is historically inaccurate. Once Boltzmann discovered the H-theorem it remained front and center in his mind. He always believed that some systems did not experience minus-H increase and he was in possession of reasons for delimiting the second law to a statistical claim well before the publication of Loschmidt's reversibility objection in 1876. But even after wrestling with that objection, Boltzmann always remained pragmatically committed to the project of mechanically justifying the second law. It is therefore in a truly Boltzmannian spirit that I have tried to resolve the reversibility paradox in a way that remains true to mechanical natural philosophy.

There remains at least one puzzle to solve. How ought the probabilities in the proposed causal Boltzmannian approach to be interpreted? I've shown that both Maxwell and Boltzmann favored (at least at one time) epistemic interpretations of the involved probabilities, and I believe that is the best option in this context. Of course, a lot more needs to be said about these epistemic probabilities, but I hope to articulate my opinions about the matter in a part two essay that uses the framework of this project to tackle the famous recurrence objections.

## Appendix 1: The Second Law of Thermodynamics in Boltzmannian Statistical Mechanics

(**The Second Law of Thermodynamics (SL)**): Necessarily, [with respect to "an arbitrary instant $t = t_1$" and a statistical mechanical system (SYS) at $t_1$, if SYS's "Boltzmann entropy…at that time, $S_B(t_1)$, is far below its maximum value", it will be "highly probable that at any later time $t_2$" ($t_2 > t_1$), "we have $S_B(t_2) > S_B(t_1)$"] and necessarily, [if SYS is at an arbitrary time $t_1$ in thermal equilibrium, then it will be "highly probable that at any later time $t_2$" ($t_2 > t_1$) we have $S_B(t_2) = S_B(t_1)$].[207]

## Appendix 2: Lanford's Project and the Chaos Asymmetry Problem[208]

Oscar Lanford III realized that in order to solve what I have called the **Chaos Asymmetry Problem** (**CAP**) he needed a **Hypothesis of Molecular Chaos** (**HMC**) that outstrips the factorization condition used in his result. Thus, I believe that Lanford's work supports the view that the **HMC** is not represented by the factorization condition needed for his famous theorem. This supports my judgment that there really is a **No Mathematics Problem** (**NMP**).

Consider:

When Lanford derived the Boltzmann equation from classical Hamiltonian mechanics for the Boltzmann-Grad limit and for a rarefied gas approximated by hard spheres, he assumed a factorization condition not unlike that which is stated in footnote 175.[209] However, Lanford perceived that there was something more lurking beneath his *time-reversal invariant* theorem that supports the time-*asymmetric* Boltzmann equation and helps represent irreversible entropic increase or equilibration governed by the inequality: $\frac{dH}{dt} \leq 0$. We witness irreversible evolutions. We measure non-equilibrium systems and reliably track their march toward equilibrium over time. To save the phenomena, we have to ensure that we secure and use the Boltzmann equation and not the anti-Boltzmann equation (which is the Boltzmann equation with the sign of the relevant collision integral flipped). These two equations are demonstrably inequivalent (Uffink and Valente 2010, 167-168). To acquire the Boltzmann equation, one can use Lanford's theorem, but *one must assume* that collision point configurations are incoming and not outgoing (ibid.). Incoming configurations determine a positive collision operator, while outgoing configurations yield the same operator with its sign flipped. It has been shown that even if one applies the time-reversal operation to incoming configurations or representations one does not obtain configurations equivalent to outgoing configurations (ibid., 172, proposition 5). Thus, there is something deeply irreversible obtained by Lanford's project and the factorization condition is insensitive to it because that condition says nothing about which set of representations or configurations one should choose. The factorization condition works equally well with incoming *or* outgoing collision phase point representations (Lanford 1975, 88). That is why Lanford himself "consistently stressed that mere factorization is not in itself the explanation of irreversibility."[210] And that is why Lanford maintained that the:

> …inequality $\frac{dH}{dt} \leq 0$ shows that the reversibility of the underlying molecular dynamics has been lost in passing to the Boltzmann equation. The irreversibility must have been introduced in the Hypothesis of Molecular Chaos since the rest of the derivation was straightforward mechanics. Indeed, it is not hard to see directly that the Hypothesis of Molecular Chaos is asymmetric in time…One conclusion which must be drawn is that something more is involved in the Hypothesis of Molecular Chaos than simple statistical independence.[211]

[207] (Frigg 2008, 105). I have changed Frigg's inequality from greater than or equal to, to just greater than.

[208] Here I'm in broad agreement and am indebted to (Uffink and Valente 2010).

[209] Again, for a precise statement of Lanford's theorem, see (Spohn 1991, 64 theorem 4.5). Spohn also provides a rigorous statement of the necessary factorization condition.

The **HMC** was something beyond the factorization condition, for the factorization condition is itself time-symmetric.

If one focuses on the beautiful mathematical result that is Lanford's theorem alone one will be unable to save the phenomenon that is irreversible thermodynamic system evolution even if in the appropriate limit. For Lanford, the closest mathematical model of what we seek to save comes not from his theorem but from Boltzmann's.

> None of this [Lanford's theorem etc.], however, really implies that irreversible behavior *must* occur in the limiting regime; it merely makes this behavior plausible. For a really compelling argument in favor of irreversibility, it seems to be necessary to rely on some version of Boltzmann's original proof of the H-theorem.[212]

But as I noted, Boltzmann's H-theorem requires the **HMC** as I have presented it. Thus, we may conjoin to the conclusion that (a) Lanford's project remains burdened by the **No Mathematics problem** the further conclusion that (b) it cannot meet the **Chaos Asymmetry Problem**. Uffink and Valente (2010) (and it seems Lanford (1975), (1981)) agree with (b), while agreement with (a) can be found in (Villani 2002) and perhaps (Spohn 1991, 76).

---

[210] (Uffink and Valente 2010, 160).
[211] (Lanford 1975, 81).
[212] (Lanford 1981, 75) emphasis in the original.

**Abbreviations:**

*BWAn* *Wissenschaftliche Abhandlungen von Ludwig Boltzmann*, edited by Fritz Hasenöhrl (Leipzig: Barth, 1909), vol. *n*.

*SPMn* *The Scientific Papers of James Clerk Maxwell*, edited by W.D. Niven. (Cambridge: Cambridge University Press, 1890), vol. *n*.

## Works Cited[213]

[213] For *Annalen der Physik* or *Annalen der Physik und Chemie* (the latter title was used from 1824 to 1899), I cite volume numbers in accord with the norms established by the journal in June of 2010.